\documentclass[aps,prb,groupedaddress]{revtex4}
\usepackage{graphicx}
\usepackage{epstopdf}
\usepackage{feynmf}	
\usepackage{sparticles}
\usepackage[margin=10pt,font=small,labelfont=bf]{caption}

\begin{document}

\title{Dark Matter: A Primer}

\author{Katherine Garrett} \email{KatherineGarrett@creighton.edu}
\author{Gintaras D$\bar \textrm{u}$da} \email{gkduda@creighton.edu}
\affiliation{Department of
Physics, Creighton University, 2500 California Plaza, Omaha, NE
68178, USA}

\begin{abstract}
  \vbox{Dark matter is one of the greatest unsolved mysteries in
  cosmology at the present time.  About 80\% of the   universe's
  gravitating matter is non-luminous, and its nature and
  distribution   are for the most part unknown.  In this paper,
  we will outline the   history, astrophysical evidence, candidates,
  and detection methods of dark matter, with the goal to give the
  reader an accessible but rigorous introduction to the puzzle of
  dark matter.  This review targets advanced students and researchers new to the field of dark 
  matter, and includes an extensive list of references for further study.}
\end{abstract}

\maketitle

\section{Introduction}

One of the most astounding revelations of the twentieth century in
terms of our understanding of the Universe is that ordinary
baryonic matter, that is, matter made up of protons and neutrons,
is not the dominant form of material in the Universe.  Rather, some
strange new form of matter, dubbed ``dark matter," fills our
Universe, and it is roughly five times more abundant than ordinary matter.
Although we have yet to detect this strange material in the laboratory,
there is a great deal of evidence which points to the necessity of its
existence.  

A complete understanding of dark matter requires utilizing several
branches of physics and astronomy.  The creation of dark matter during
the hot expansion of the universe is understood through
statistical mechanics and thermodynamics.  Particle physics is
necessary to propose candidates for dark matter and explore its
possible interactions with ordinary matter.  General relativity,
astrophysics, and cosmology dictate how dark matter acts on large
scales and how the universe may be viewed as a laboratory to
study dark matter.  Many other areas of physics come into play as
well, making the study of dark matter a diverse and
interdisciplinary field.  Furthermore, the profusion of ground and satellite-based measurements
in recent years have rapidly advanced the field making it dynamic and timely; we are truly entering the era of ``precision cosmology".

This paper aims to give a general overview of the subject of dark matter
suitable for non-experts; we hope to treat this fascinating and important topic in a way
such that the non-specialist will gain a strong foundation and
introduction to dark matter.  It is at times difficult to find
understandable and appropriate literature for individuals with no
background on the subject.  Existing reviews are either popular-level pieces which
are too general or specialized pieces for experts in the field, motivating us to create an accessible overview.  We particularly hope that this review will be helpful to graduate students beginning their study of dark matter and to other physicists and astronomers who would like to learn more about this important topic.  

To give such an introduction to dark matter, we will first briefly explain the
first hints that dark matter exists, elaborate on the strong
evidence physicists and astronomers have accumulated in the past
years, discuss the neutralino and other possible candidates, and
describe various detection methods used to probe the dark matter's
mysterious properties.  Although we will at times focus on supersymmetric theories of dark matter, other possibilities will be introduced and discussed.

\section{History and Early Indications}

Astronomers have long relied on photometry to yield estimates on
mass, specifically through well defined mass to luminosity ratios
($M/L$). This is not at all surprising, since visual astronomy
relies on the light emitted from distant objects.  For example,
the $M/L$ ratio for the sun is $M/L = 5.1 \times
10^3~\textrm{kg/W}$; since this number is not terribly
instructive, one usually measures mass to luminosity in terms of
the sun's mass and luminosity such that $M_\odot/L_\odot = 1$ by
definition.  Thus by measuring the light output of an object (for
example a galaxy or cluster of galaxies) one can use well-defined
$M/L$ ratios in order to estimate the mass of the object.

In the early 1930s, J. H. Oort found that the motion of stars in
the Milky Way hinted at the presence of far more galactic mass
than anyone had previously predicted. By studying the Doppler
shifts of stars moving near the galactic plane, Oort was able to
calculate their velocities, and thus made the startling discovery
that the stars should be moving quickly enough to escape the
gravitational pull of the luminous mass in the galaxy. Oort
postulated that there must be more mass present within the Milky
Way to hold these stars in their observed orbits.  However, Oort
noted that another possible explanation was that 85\% of the light
from the galactic center was obscured by dust and intervening
matter or that the velocity measurements for the stars in question
were simply in error.\cite{Oort}

Around the same time Oort made his discovery, Swiss astronomer
F. Zwicky found similar indications of missing mass, but on a much
larger scale.  Zwicky studied the Coma cluster, about 99 Mpc (322 million
lightyears) from Earth, and, using observed doppler shifts in
galactic spectra, was able to calculate the velocity dispersion of
the galaxies in the Coma cluster. Knowing the velocity dispersions
of the individual galaxies (i.e. kinetic energy), Zwicky employed
the virial theorem to calculate the cluster's mass.  Assuming only
gravitational interactions and Newtonian gravity ($F \propto 1/r^2$),
the virial theorem gives the following relation
between kinetic and potential energy:

\begin{equation}
\langle T \rangle= -\frac{1}{2} \langle U \rangle, \label{virial theorem2}
\end{equation}

\noindent where $\langle T \rangle$ is the average kinetic energy and
$\langle U \rangle$ is the average potential energy.  Zwicky found that the total 
mass of the cluster was $M_{cluster} \approx 4.5 \times 10^{13} M_{\odot}$).  Since he
observed roughly 1000 nebulae in the cluster, Zwicky calculated that the average mass of
each nebula was $M_{nebula} = 4.5 \times 10^{10} M_{\odot}$.
This result was startling because a measurement of the mass of the cluster
using standard $M/L$ ratios for nebulae gave a total mass for the cluster approximately
2\% of this value. In essence, galaxies only accounted for only a small fraction of the total mass; the vast majority of the mass of the Coma
cluster was for some reason ``missing" or non-luminous (although not known to Zwicky at the time,
roughly 10\% of the cluster mass is contained in the intracluster gas which slightly alleviates but does not solve the issue of missing mass).\cite{Zwicky1,Zwicky2}

Roughly 40 years following the discoveries of Oort, Zwicky, and others, Vera Rubin
and collaborators conducted an extensive study of the rotation
curves of 60 isolated galaxies.\cite{Rubin}  The galaxies chosen
were oriented in such a way so that material on one side of the
galactic nucleus was approaching our galaxy while material on the
other side was receding; thus the analysis of spectral lines
(Doppler shift) gave the rotational velocity of regions of the
target galaxy.  Additionally, the position along the spectral line
gave angular information about the distance of the point from the
center of the galaxy.   Ideally one would target individual stars
to determine their rotational velocities; however, individual
stars in distant galaxies are simply too faint, so Rubin used
clouds of gas rich in hydrogen and helium that surround hot stars
as tracers of the rotational profile.

It was assumed that the orbits of stars within a galaxy would
closely mimic the rotations of the planets within our solar
system. Within the solar system,

\begin{equation}
v(r) = \sqrt{G \frac{m(r)}{r}},
\label{kepler's law}
\end{equation}

\noindent where $v(r)$ is the rotation speed of the object at a
radius $r$, $G$ is the gravitational constant, and $m(r)$ is the
total mass contained within $r$ (for the solar system essentially the sun's mass),
which is derived from simply
setting the gravitational force equal to the centripetal force
(planetary orbits being roughly circular). Therefore, $v(r) \propto
1/ \sqrt{r}$, meaning that the velocity of a rotating body should
decrease as its distance from the center increases, which is
generally referred to as ``Keplerian" behavior.

\begin{figure}[htp]
\includegraphics[width=0.50\textwidth]{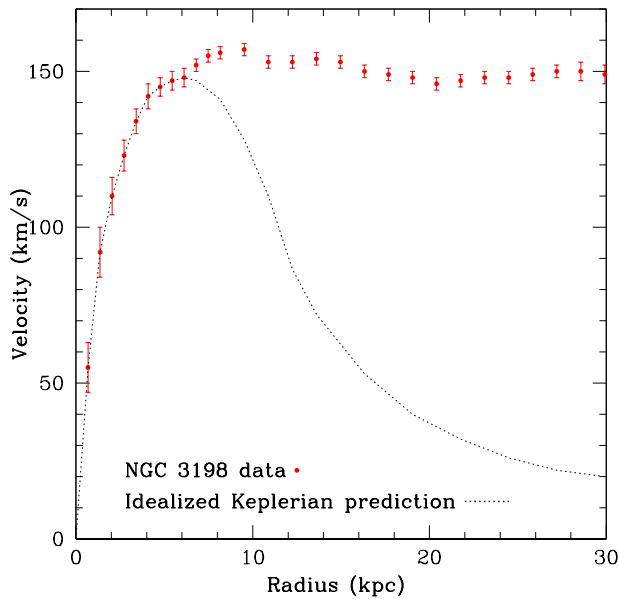}
\caption{Measured rotational velocities of HI regions in NGC 3198 [5] compared to an idealized Keplerian behavior.} \label{fig:1}
\end{figure}

Rubin's results showed an extreme deviation from predictions due
to Newtonian gravity and the luminous matter distribution. The
collected data showed that the rotation curves for stars are
``flat," that is, the velocities of stars continue to
\emph{increase} with distance from the galactic center until they
reach a limit (shown in Fig.~\ref{fig:1}).  An intuitive way to
understand this result is through a simplified model: consider the galaxy as a uniform sphere of mass and apply Gauss's law for gravity (in direct analogy with Gauss's Law for the electric field):
\begin{equation}
\int_S \vec g \cdot d\vec A = 4 \pi G M_{encl}, \label{gauss's law
for gravity} \end{equation}

\noindent where the left hand side is the flux of the
gravitational field through a closed surface and the right hand
side is proportional to the total mass enclosed by that surface.
If, as the radius of the Gaussian surface increases, more and more
mass in enclosed, then the gravitational field will grow; here velocities can grow or remain
constant as a function of radius $r$ (with the exact behavior depending on the mass profile $M(r)$).
If, however, the mass enclosed decreases or remains constant as the Gaussian surface grows, then
the gravitational field will fall, leading to smaller and smaller
rotational velocities as $r$ increases.  Near the center of the galaxy where the
luminous mass is concentrated falls under the former condition,
whereas in the outskirts of the galaxy where little to no
additional mass is being added (the majority of the galaxy's mass
being in the central bulge) one expects the situation to be that
of  the latter.  Therefore, if the rotational velocities remain
constant with increasing radius, the mass interior to this radius
must be increasing.  Since the density of luminous mass falls past
the central bulge of the galaxy, the ``missing" mass must be
non-luminous.  Rubin summarized, ``The conclusion is inescapable:
mass, unlike luminosity, is not concentrated near the center of
spiral galaxies.  Thus the light distribution in a galaxy is not
at all a guide to mass distribution."\cite{Rubin}

In the 1970s, another way to probe the amount and distribution of
dark matter was discovered: gravitational lensing. Gravitational
lensing is a result of Einstein's Theory of Relativity which postulates
that the universe exists within a flexible fabric of spacetime.
Objects with mass bend this fabric, affecting the motions of bodies
around them (objects follow geodesics on this curved surface). The
motions of planets around the sun can be explained in this way, much
like how water molecules circle an empty drain. The path of light is
similarly affected; light bends when encountering massive objects.
To see the effects of gravitational lensing, cosmologists look for a relatively
close, massive object (often a cluster of galaxies) behind which a
distant, bright object (often a galaxy) is located (there is actual an optimal lens-observer separation, so this must be taken into account as well). If the distant
galaxy were to be located \emph{directly} behind the cluster, a
complete ``Einstein ring" would appear; this looks much like a bullseye,
where the center is the closer object and the ring is the lensed
image of the more distant object.  However, the likelihood of
two appropriately bright and distant objects lining up perfectly
with the Earth is low; thus, distorted galaxies generally appear
as ``arclets," or partial Einstein rings.

In 1979, D. Walsh \emph{et al.} were the first to observe
this form of gravitational lensing.  Working at the Kitt Peak National Observatory,
they found two distant objects separated by only 5.6 arc seconds with
very similar redshifts, magnitudes, and spectra.\cite{Walsh} They concluded
that perhaps they were seeing the same object twice, due to the lensing of
a closer, massive object. Similar observations were made by R. Lynds and
V. Petrosian in 1988, in which they saw multiple arclets within clusters.\cite{Lynds}

We can study a distant galaxy's distorted image and make conclusions
about the amount of mass within a lensing cluster using this expression for
$\theta_E$, the ``Einstein radius" (the radius of an arclet in radians):

\begin{equation}
\theta_E = \sqrt{\frac{4GM}{c^2}\frac{d_{LS}}{d_L d_S}}
\end{equation}

\noindent where $G$ is the gravitational constant, $M$ is the mass
of the lens, $c$ is the speed of light,and $d_{LS}$, $d_L$, and
$d_S$ are the distance between the lens and source, the distance
to the lens, and the distance to the source, respectively (note: these distances are angular-diameter distances which differ from our ``ordinarily" notion of distance, called the proper distance, due to 
the expansion and curvature of the universe).
Physicists have found that this calculated mass is much larger
than the mass that can be inferred from a cluster's luminosity.
For example, for the lensing cluster Abell 370, Bergmann,
Petrosian, and Lynds determined that the $M/L$ ratio of
the cluster must be about 10$^2$-10$^3$ solar units, necessitating the
existence of large amounts of dark matter in the cluster as well as
placing constraints on its distribution within the cluster.\cite{BPL}

\section{Modern Understanding and Evidence}

\subsection{Microlensing}

To explain dark matter physicists first turned to astrophysical objects made of ordinary, baryonic matter (the
type of matter that we see everyday and is made up of fundamental
particles called quarks, which we will discuss in further detail
in section IV).  Since we know that dark matter must be ``dark,"
possible candidates included brown dwarfs, neutron
stars, black holes, and unassociated planets; all of these candidates
can be classified as MACHOs (MAssive Compact Halo Objects).

To hunt for these objects two collaborations, the MACHO Collaboration
and the EROS-2 Survey, searched for gravitational microlensing
(the changing brightness of a distant object due to the interference
of a nearby object) caused by possible MACHOs in the Milky Way halo. (Other
collaborations have studied this as well, such as MOA, OGLE, and SuperMACHO.\cite{MOA,OGLE,SuperMACHO})
The MACHO Collaboration painstakingly observed and statistically analyzed
the skies for such lensing; 11.9 million stars were studied, with only 13-17
possible lensing events detected.\cite{MACHO}  In April of 2007, the
EROS-2 Survey reported even fewer events, observing a sample of 7
million bright stars with only \emph{one} lensing candidate found.\cite{EROS2}
This low number of possible MACHOs can only account for a very small
percentage of the non-luminous mass in our galaxy, revealing that most dark matter
cannot be strongly concentrated or exist in the form of baryonic astrophysical objects.
Although microlensing surveys rule out baryonic objects like brown dwarfs, black holes, and neutron stars in our galactic halo, can other forms of baryonic matter make up the bulk of dark matter?  The answer, surprisingly, is no, and the evidence behind this claim comes from Big Bang Nucleosynthesis (BBN) and
the Cosmic Microwave Background (CMB).

\subsection{Cosmological Evidence}

\begin{figure}[htp]
\includegraphics[width=0.50\textwidth]{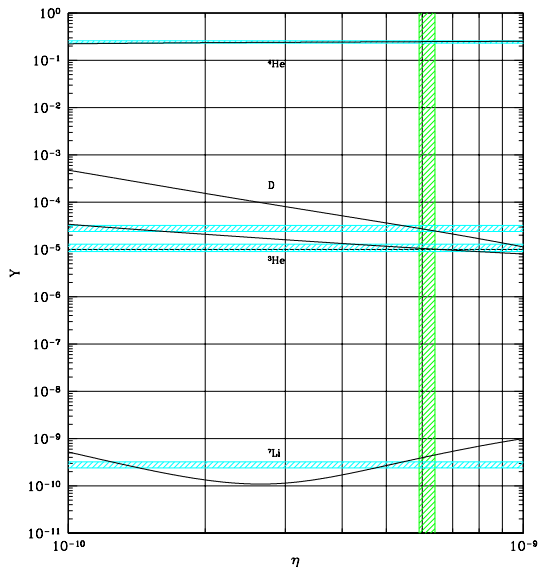}
\caption{Light elemental abundances versus the photon to baryon ratio, $\eta$.  The horizontal lines show measured abundances of the respective elements and the vertical lines show the photon to baryon ratio as measured by WMAP.} \label{fig:BBN}
\end{figure}

BBN is a period from a few seconds to a few minutes after the Big Bang in the early, hot universe when neutrons and protons fused together to form deuterium, helium, and trace amounts of lithium and other light elements.  In fact, BBN is the largest source of deuterium in the universe as any deuterium found or produced in stars is almost immediately destroyed  (by fusing it into $^4$He); thus the present abundance of deuterium in the universe can be considered a ``lower limit" on the amount of deuterium created by the Big Bang.    Therefore, by considering the deuterium
to hydrogen ratio of distant, primordial-like areas with low levels of
elements heavier than lithium (an indication that these areas have not
changed significantly since the Big Bang), physicists are able to
estimate the D/H abundance directly after BBN (it is useful to look at the ratio of a particular element's abundance relative to hydrogen).  Using nuclear physics and known
reaction rates, BBN elemental abundances can be theoretically calculated; one
of the triumphs of the Big Bang model is the precise agreement between
theory and observational determinations of these light elemental
abundances.   Fig.~(\ref{fig:BBN}) shows theoretical elemental abundances as calculated with the BBN code nuc123 compared with experimental ranges.\cite{nuc123}  It turns out that the D/H ratio is heavily dependent on the overall density of baryons in the universe, so measuring the D/H
abundance gives the overall baryon abundance.  This is usually
represented by $\Omega_b h^2$, where $\Omega_b$ is the baryon density
relative to a reference critical density ($\rho_c$) and $h=H/100~\textrm{km sec}^{-1}~\textrm{Mpc}^{-1}$ (the reduced Hubble constant, which is used because of the large historical uncertainty in the expansion rate of the universe).  R. H. Cyburt calculated two possible values for $\Omega_b
h^2$ depending on what deuterium observation is taken: $\Omega_b h^2
= 0.0229 \pm 0.0013$ and $\Omega_b h^2 = 0.0216_{-0.0021}^{+0.0020}$,
both which we will see accounts for only about 20\% of the total
matter density.\cite{Cyburt}

The CMB, discovered by Penzias and Wilson
in 1964 (but theorized by others much earlier) as an excess background
temperature of about 2.73 K, is another way in which we can
learn about the composition of the universe.\cite{Penzias}  Immediately
after the Big Bang, the universe was an extremely dense plasma of charged
particles and photons.  This plasma went through an initial rapid
expansion, then expanded at a slower, decreasing rate, and cooled for about
380,000 years until it reached what is known as the epoch of recombination.
At this time, neutral atoms were formed, and the universe became transparent to electromagnetic radiation; in other words, photons, once locked to charged particles because of interactions, were now able to travel unimpeded through the universe.  The photons released from this ``last scattering" exist today as the CMB.

COBE (COsmic Background Explorer) launched in 1989, verified two fundamental properties of the CMB: 1) the CMB is remarkably uniform (2.73 K across the sky) and 2) the CMB, and thus the early universe, is a nearly perfect blackbody (vindicating the use of statistical thermodynamics to describe the early universe).  Although the CMB is extraordinarily uniform,
COBE's Differential Microwave Radiometer (DMR) discovered in its first year fundamental anisotropies
(fluctuations) within the CMB, beyond the signal due to our motion relative to the CMB frame and foregrounds, such as emission from dust in the Milky Way.   These fundamental fluctuations are due to two
different effects. Large scale fluctuations can be attributed to the
Sachs-Wolfe effect: lower energy photons are observed today from areas that were more dense at the time of last scattering (these photons, once emitted, lost energy escaping from deeper gravitational potential wells).
On small scales, the origin of the CMB anisotropies are due to what are
called acoustic oscillations. Before photon decoupling, protons
and photons can be modeled as a photon-baryon fluid (since electrons are so much less
massive than baryons we can effectively ignore them here). This fluid effectively goes through the following cycle: 1) the fluid is compressed as it falls into a gravitational well, 2) the pressure of
the fluid increases until it forces the fluid to expand outward, 3)
the pressure of the fluid decreases as it expands until gravity pulls
it back, and 4) the process repeats
until photon decoupling.  Depending on the location in the cycle for a
portion of the fluid at photon decoupling, the photons which emerge
vary in temperature. The fluctuations in the CMB are thus indications of both the initial density perturbations that allowed for the formation of early gravitational wells as well as dynamics of the photon-baryon fluid.  In this manner the temperature fluctuations of the CMB are dependent on the amount of baryons in the universe at the time of recombination.

Although the detection of the fluctuations in the CMB was a major accomplishment, the magnitude
of the temperature variations puzzled scientists.   These fundamental fluctuations in the CMB are incredibly small, only about $30 \pm 5$
$\mu$K, meaning that the CMB is uniform to 1 part in $10^5$.  In fact, these fluctuations were too small to have solely accounted for the seeds of structure
formation;\cite{COBE} essentially, given the size of the CMB
fluctuations, the structure of the universe we see today would not
have had time to form.  The problem is time: ordinary matter only
becomes charge neutral at the epoch of recombination, and before that,
due to electrostatic forces, matter cannot effectively clump into
gravitational wells to begin forming structure.  The COBE results
showed a need for an electrically neutral form of matter that could
jump start the structure formation process well before recombination.

\begin{figure}[htp]
\includegraphics[width=0.50\textwidth]{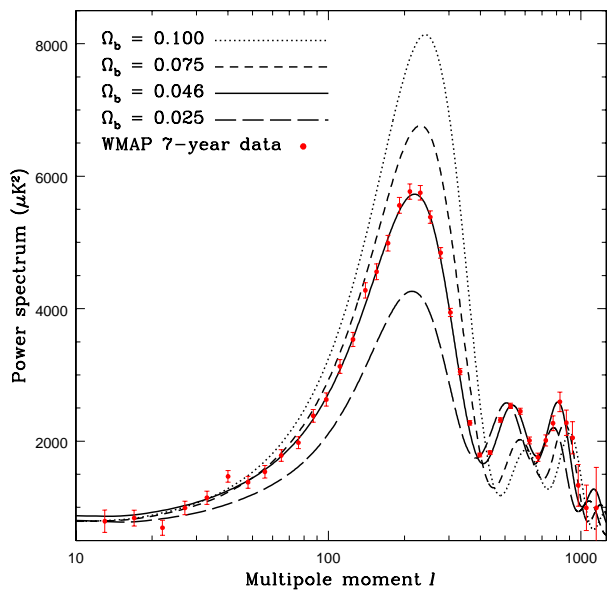}
\caption{The CMB Anisotropy Power Spectrum for various values of $\Omega_b$ and $\Omega_{dm}$ (holding $\Omega_{tot}=1$) with WMAP year 7 data.   The anisotropy power spectrum gives the level of temperature fluctuations on patches of various angular scales, where a spherical version of a Fourier transform gives multipoles $l$, where roughly $l = 180^\circ /\theta$, with $\theta$ the angular scale in degrees.} \label{fig:cmb}
\end{figure}

WMAP (Wilkinson Microwave Anisotropy Probe) was launched in 2001 with the mission to more precisely measure the anisotropies in the CMB.  Located at the Earth-Sun L2 point (about a million miles from Earth), the satellite has taken data continuously (most recently having released an analysis of seven years of operation)  and is able to detect temperature variations as small as one millionth of a degree.  Due to the increased angular resolution of WMAP (and through the use of computer codes which can calculate the CMB anisotropies given fundamental parameters such as the baryon density) we now know the total and baryonic matter densities from WMAP:\cite{WMAP}

\begin{equation}
\Omega_m h^2 = 0.1334^{+0.0056}_{-0.0055}, ~~~ \Omega_b h^2 = 0.02260 \pm 0.00053,
\end{equation}

\noindent where $\Omega_m h^2$ is the total matter density, and
$\Omega_b h^2$ is the baryonic matter density.  The first essential
observation is that these two numbers are different; baryonic matter
is not the only form of matter in the universe.  In fact, the dark matter density,
$\Omega_{dm} h^2 = 0.1123 \pm 0.0035$,  is around 83\% of the total mass density.  Locally, this corresponds to an average density of dark matter $\rho_{dm} \approx 0.3~\textrm{GeV/cm}^3 \approx 5 \times 10^{-28}~\textrm{kg/m}^3$ at the Sun's location (which enhanced by a factor of roughly $10^5$ compared to the overall dark matter density in the universe due to structure formation).   An analysis of the CMB allows for a discrimination between dark matter and ordinary matter precisely because the two components act differently; the dark matter accounts for roughly 85\% of the mass, but unlike the baryons, it is not linked to the photons as part of the ``photon-baryon fluid."  Fig.~(\ref{fig:cmb}) demonstrates this point extremely well; small shifts in the baryon density result in a CMB anisotropy power spectrum (a graphical method of depicting the  CMB anisotropies) which are wholly inconsistent with WMAP and other CMB experiment data.

Analyses of the large scale structure of the universe also yield evidence
for dark matter and help break degeneracies present in the CMB data analysis.
By calculating the distance to galaxies using their redshifts,
cosmologists have been able to map out the approximate locations of
more than 1.5 million galaxies.  For example, the Sloan Digital Sky Survey
(SDSS) has created 3-D maps of more than 900,000 galaxies, 120,000 quasars,
and 400,000 stars during its eight years of operation.\cite{SDSS}  In fact,
galaxy counts have had a long and important history in
cosmology; in the 1950s and 60s radio galaxy counts provided the earliest,
hard evidence against the Steady State model.  But how can galaxy counts give evidence for dark matter?  As discussed earlier, the current structure in the universe is
due to initial density fluctuations which served as seeds for structure formation magnified by the presence of dark matter.  The most likely source of these initial density perturbations are quantum fluctuations magnified by inflation, a period of early rapid exponential growth approximately $10^{-35}$ seconds after the big bang.  Under the assumption that these random fluctuations are Gaussian, a single function, the power spectrum $P(k$), is sufficient to describe the density perturbations.   From here a given $P(k)$ can be used to theoretically calculate large scale structure.  These statements are true, of course, only statistically.  Furthermore, the converse is also true: by measuring large scale structure (galaxy counts and surveys) one can experimentally determine the power spectrum $P(k)$.   By obtaining the matter power spectrum from galaxy surveys, the amount of total matter and baryonic
matter can be found: the peak of $P(k)$ is sensitive to the value
of $\Omega_m$, and the amount of baryons has effects on the shape of $P(k)$ (through baryonic acoustic oscillations, i.e. excesses in galaxies separated at certain distances due to sound waves in
the pre-recombination plasma)\cite{BAO}.
Using these techniques, a final study of the 2dF Galaxy Redshift Survey power spectrum found
$\Omega_m=0.231\pm0.021$ and $\Omega_b/\Omega_m=0.185\pm0.046$; a study based on
data from SDSS yielded $\Omega_m=0.286\pm0.018$ and $\Omega_{dm} h^2=0.02267\pm0.00058$.\cite{2DF,SDSS2}  Note that these results agree with both CMB and BBN predictions.

\emph{N}-body simulations of Large Scale Structure are another tool which have been used to demonstrate the need for dark matter.  These simulations often take weeks to complete on superclusters; for example, MS-II tracked over 10 billion particles which each represent 6.89$\times$10$^6~h^{-1}~M_\odot$ in a volume of (100 $h^{-1}$ Mpc)$^3$ to study dark matter halo structure and formation.\cite{Boylan-Kolchin} Similarly, di Matteo \emph{et al.} ran simulations to study the role of black holes in structure formation using 20-200 million particles in a volume of (33.75 $h^{-1}$ Mpc)$^3$ to (50 $h^{-1}$ Mpc)$^3$.\cite{di Matteo}  N-body simulations confirm the need for dark matter.
Simulations
without dark matter do not form the familiar filament and void-type structures seen in the observable universe by SDSS and other surveys on the proper timescales.  Additionally,
scenarios run in which dark matter is relativistic
or ``hot" find that structure formation is retarded or ``washed-out" instead of enhanced; thus not only is dark matter needed, but more specifically, dark matter must be ``cold"  or nonrelativistic during
the period of structure formation.\cite{Peebles,Jenkins}

\subsection{Most recent evidence}

Recent evidence hailed as the ``smoking-gun" for dark matter comes from the Bullet cluster,
the result of a sub-cluster (the ``bullet") colliding with the larger
galaxy cluster 1E 0657-56.  During the collision, the galaxies within the two clusters
passed by each other without interacting (a typical
distance between galaxies is approximately one megaparsec, or 3.26 million
lightyears).  However, the majority of a cluster's baryonic mass exists in the extremely hot gas between
galaxies, and the cluster collision (at roughly six million miles per hour) compressed and shock heated this gas; as a result, a
huge amount of X-ray radiation was emitted which has been observed by NASA's
Chandra X-ray Observatory.  Comparing the location of this radiation
(an indication of the location of the majority of the baryonic mass in the clusters) to a mapping of
weak gravitational lensing (an indication of the location of the majority of the total mass  of the clusters) shows an interesting discrepancy; the areas of strong
X-ray emission and the largest concentrations of mass seen through gravitational lensing are not the same. The majority of the mass in the clusters is
non-baryonic and gravity ``points" back to this missing mass.\cite{Clowe}

The galaxy cluster known as MACS J0025.4-1222 is a second example of
a powerful collision between two clusters which separated the luminous
and dark matter within the two clusters. In mid-2008, Brada$\check
\textrm{c}$ \emph{et al.} found that the behavior of the matter within
this cluster is strikingly similar to the Bullet Cluster; the dark
matter passed through the collision while the intergalactic gas interacted
and emitted X-rays. These results reaffirmed those of the Bullet Cluster
and the need for collisionless dark matter (as well as severely constraining MOND theories (see the Appendix for a brief description of MOND theories).\cite{Bradac}

In May of 2007, NASA's Hubble Space Telescope (HST) detected a ring-like
structure of dark matter, caused by another collision of two massive
galaxy clusters one to two billion years ago.\cite{Jee}  The dark
matter in the two clusters collapsed towards the center, but some of
it began to ``slosh" back out, causing the ring-shaped structure it
now has. Overlapping the distribution of gravitational lensing with
the baryonic mass in the combined cluster (just as with the Bullet
cluster) shows the largest discrepancy yet between luminous and dark
matter.

In early 2009, Penny \emph{et al.} released results of a study from a
HST survey of the Perseus Cluster, which is located about 250 million
light years from Earth. They noticed that small, dwarf spheroidal galaxies
are stable while larger galaxies are being torn apart by tidal
forces caused by the cluster potential, a sign that a significant amount of
dark matter may be holding the dwarf galaxies together.  By determining the minimum
mass required for these dwarf galaxies to survive the cluster potential, Penny
\emph{et al.} were able to calculate the amount of dark matter needed in
each galaxy. They specifically studied 25 dwarf spheroidal galaxies
within the cluster and found that 12 require dark matter in order to survive
the tidal forces at the cluster center.

In conclusion, the evidence for dark matter on scales from dwarf galaxies to clusters to the largest scales in the universe
is compelling.  There is remarkable agreement between multiple lines of evidence about the need for cold dark matter.  Having established the need for dark
matter, in the next section we will discuss possible particle candidates for dark matter and how theories beyond the Standard Model are necessary to solve the puzzle.

\section{Particle Candidates}
Although the existence of dark matter is well motivated by several lines of evidence,
the exact nature of dark matter remains elusive.  Dark matter candidates are generically
referred to as WIMPs (Weakly Interacting Massive Particles); in other words, they are massive particles that are electrically neutral which do not interact very strongly with other matter.  In this section we will explore some possible particle candidates for dark matter and the theories that lie behind them.  But to begin with, we give a brief review of the Standard Model of particle physics.

\subsection{The Standard Model and the Neutrino}
The Standard Model (SM) is the quantum field theory that describes three of the four fundamental forces in nature: electricity and magnetism, the weak nuclear force, and the strong nuclear force.   Gravitational interactions are not part of the SM; at energies below the Planck scale gravity is unimportant at the atomic level.  There are sixteen confirmed particles in the SM, seven of which were
predicted by the model before they were found experimentally, and one
particle yet to be seen:  the Higgs boson, which is believed to be
the mediator of the Higgs field responsible for giving all other SM
particles mass.  In the SM, there are six quarks (up, down, top,
bottom, charm, and strange), six leptons (electron, mu, tau, and their
respective neutrinos), and five force carriers (photons, gluons,
$W^{\pm}$, $Z$, and the Higgs boson).   Quarks and leptons are classified as fermions with
half integer spins and are split into three generations, where force
carriers are classified as gauge bosons with integer spins.  Each of
these particles also has a corresponding antiparticle, denoted with
a bar (for example, the up antiquark's symbol is $\bar{u}$), with
opposite charge.  Table~\ref{tab:1} arranges the SM fundamental
particles and some of their basic qualities.

\begin{table}[h]
\begin{center}

\begin{tabular}{l}
\textbf{Fermions}
\end{tabular}

\begin{tabular}{|lcc|lcc|lcc|}
\hline ~Generation 1 &&& ~Generation 2 &&& ~Generation 3 &&\\
\hline ~Particle & Mass (MeV) & ~Charge~ & ~Particle & Mass (MeV) & ~Charge~ & ~Particle & Mass (MeV) & ~Charge~ \\
\hline ~up quark ($u$) & 2.55 & +$\frac{2}{3}$ & ~charm quark ($c$) & 1270 & +$\frac{2}{3}$ & ~top quark ($t$) & 171200 & +$\frac{2}{3}$\\
~down quark ($d$) & 5.04 & $-\frac{1}{3}$ & ~strange quark ($s$) & 104 & $-\frac{1}{3}$ & ~bottom quark ($b$) & 4200 & $-\frac{1}{3}$\\
~electron ($e^-$) & 0.511 & $-$1 & ~muon ($\mu^-$) & 105.7 & $-$1 & ~tau ($\tau^-$) & 1776.8 & $-$1\\
~$e$ neutrino ($\nu_e$) ~ & $< 2.0 \times 10^{-6}$ & 0 & ~$\mu$ neutrino ($\nu_{\mu}$) & $<0.19$ & 0 & ~$\tau$ neutrino ($\nu_{\tau}$) & $< 18.2$ & 0\\
\hline
\end{tabular}

\begin{tabular}{l}
~ \\
\textbf{Gauge Bosons}
\end{tabular}

\begin{tabular}{|llllcc|}
\hline ~Particle & Force & Acts through & Acts on & Mass (MeV) & ~~Charge~~ \\
\hline ~Photon ($\gamma$) & Electromagnetic~~~~~ & Electric charge~~~ & Electrically charged particles ~~~~ & $<$ 1$\times 10^{-24} \approx 0$ &0\\
       ~$Z$ boson ($Z$) & Weak nuclear & Weak interaction~~~~~ & Quarks and leptons & 91188 & 0\\
       ~$W^\pm$ bosons ($W^\pm$)~~~~ & Weak nuclear & Weak interaction  & Quarks and leptons & 80398 & $\pm$1\\
       ~Gluon ($g$) & Strong nuclear & Color charge & Quarks and gluons & 0 & 0\\
       ~Higgs boson ($H^0$) & Higgs force & Higgs field & Massive particles & $>$ 114400 & 0 \\
\hline
\end{tabular}

\end{center}
\caption{The particles predicted by the Standard Model.  Approximate masses of particles as last reported by the Particle Data Group.\cite{PDG}}
\label{tab:1}
\end{table}

SM particle interactions obey typical conservation of momentum and energy laws as well as conservation laws for internal gauge symmetries like conservation of charge, lepton number, etc.  The model has been thoroughly probed up to energies of $\approx$ 1 TeV, and has led to spectacular results such as the precision measurement of the anomalous magnetic moment of the electron (analogous to measuring the distance between New York and Los Angeles to the width of a human hair).

The final undiscovered particle, the Higgs boson, is thought to be extremely massive; the latest bounds from the CDF and D{\O} collaborations at the Tevatron have restricted the mass of the Higgs to two regions: 114-160 GeV and 170-185 Gev.\cite{CDF&DO} Since the Higgs boson couples very weakly to ordinary matter it is difficult to create in particle accelerators. Hopefully, the powerful Linear Hadron Collider (LHC) in Geneva, Switzerland, will confirm the existence of the Higgs boson, the final particle of the SM.

Despite its success, the SM does not contain any particle that could act as the dark matter.  The only stable, electrically neutral, and weakly interacting particles in the SM are the neutrinos.    Can the neutrinos be the missing dark matter?  Despite  having the ``undisputed virtue of being known to exist" (as put so well by Lars Bergstrom), there are two major reasons why neutrinos cannot account for all of the universe's dark matter.  First, because neutrinos are relativistic, a neutrino-dominated universe would have inhibited structure formation and caused a ``top-down" formation (larger structures forming first, eventually condensing and fragmenting to those we see today).\cite{Bond}  However, galaxies have been observed to exist less than a billion years after the big bang and, together with structure formation simulations, a ``bottom-up" formation (stars galaxies then large galaxies then clusters etc.) seems to be the most likely.\cite{Iye} Second, Spergel \emph{et al.}  ruled out neutrinos as the entire solution to missing mass using cosmological observations:  WMAP combined with large-scale structure data constrains the neutrino mass to $m_v <$ 0.23 eV, which in turn makes the cosmological density $\Omega_v h^2 <$ 0.0072.\cite{WMAP}  While neutrinos do account for a small fraction of dark matter, they clearly cannot be the only source.

The lack of a dark matter candidate does not invalidate the SM, but rather suggests that it must be extended.  Perhaps the SM is only a valid theory at low energies, and that there is new physics ``beyond the Standard Model;" that is, new theories may supplement, rather than replace, the SM.  Such new theories have already been proposed, the most promising being supersymmetry, which also yields a viable dark matter candidate called the neutralino or LSP.

\subsection{Problems of the Standard Model}

Although very successful, the SM has two flaws which hint at the need for new solutions: the hierarchy problem and the fine-tuning problem.   The hierarchy problem arises from the SM's prediction of the Higgs vacuum expectation value (vev, i.e. the average value of the field in the vacuum), which is about 246 GeV.  Theorists have predicted that at high enough energies ($\approx$ 1 TeV) the electromagnetic and weak forces act as a single unified force called the electroweak force (this has also been experimentally verified).  However at smaller energies, the single unified force breaks down into two separate forces: the electromagnetic force and the weak force.  It turns out that after this breaking, the Higgs field's lowest energy state is not zero, but the 246 GeV vacuum expectation value.  It is precisely this non-zero value that gives other particles mass through their interactions with the Higgs field.  The 246 GeV vev is at the weak scale (the typical energy of electroweak processes); however, the Planck scale (the energy at which quantum effects of gravity become strong) is around 10$^{19}$ GeV.  The basic question, then, is why is the Planck scale 10$^{16}$ times larger than the weak scale?  Is there simply a ``desert" between $10^3$ and $10^{19}$ GeV in which no new physics enters?

\begin{figure}[!htb]
\begin{center}
\begin{fmffile}{a1}
   \begin{fmfgraph*}(110,62) 
    \fmfleft{i1,i2}	
    \fmfright{o1,o2}    
    \fmflabel{$e^-$}{i1} 
    \fmflabel{$e^-$}{i2} 
    \fmflabel{$e^-$}{o1} 
    \fmflabel{$e^-$}{o2} 
    \fmf{fermion}{i1,v1,i2} 
    \fmf{fermion}{o1,v2,o2} 
    \fmf{photon,label=$\gamma$}{v1,v2} 
   \end{fmfgraph*}
\end{fmffile}
\qquad\qquad
\begin{fmffile}{b1}
   \begin{fmfgraph*}(110,62) 
    \fmfleft{i1,i2}	
    \fmfright{o1,o2}    
    \fmflabel{$e^-$}{i1} 
    \fmflabel{$e^-$}{i2} 
    \fmflabel{$e^-$}{o1} 
    \fmflabel{$e^-$}{o2} 
    \fmf{fermion}{i1,v1,i2} 
    \fmf{fermion}{o1,v4,o2} 
    \fmf{photon,label=$\gamma$}{v1,v2}
    \fmf{photon,label=$\gamma$}{v3,v4} 
    \fmf{fermion,left,tension=.3,label=$e^-$}{v2,v3}
    \fmf{fermion,left,tension=.3,label=$e^+$}{v3,v2}
   \end{fmfgraph*}
\end{fmffile}
\qquad\qquad
\begin{fmffile}{c1}
   \begin{fmfgraph*}(110,62) 
    \fmfleft{i1,i2}	
    \fmfright{o1,o2}    
    \fmflabel{$e^-$}{i1} 
    \fmflabel{$e^-$}{i2} 
    \fmflabel{$e^-$}{o1} 
    \fmflabel{$e^-$}{o2} 
    \fmf{plain}{i1,v2}
    \fmf{fermion}{v2,v3}
    \fmf{plain}{v3,v1}
    \fmf{photon,right,tension=0,label=$\gamma$}{v2,v3}
    \fmf{fermion,tension=1/3}{v1,i2}
    \fmf{fermion,tension=1/3}{o1,v4}
    \fmf{plain}{v4,v5}
    \fmf{fermion}{v5,v6}
    \fmf{plain}{v6,o2}
    \fmf{photon,label=$\gamma$}{v1,v4} 
   \end{fmfgraph*}
\end{fmffile}
\end{center}
\caption{The tree level diagram on the left represents the electron-electron scattering process to the lowest order in perturbation theory.  The two graphs on the right represent higher order processes (which can be thought of as loop corrections) and enter with higher powers of $\alpha$.}
\label{fig:4}
\end{figure}
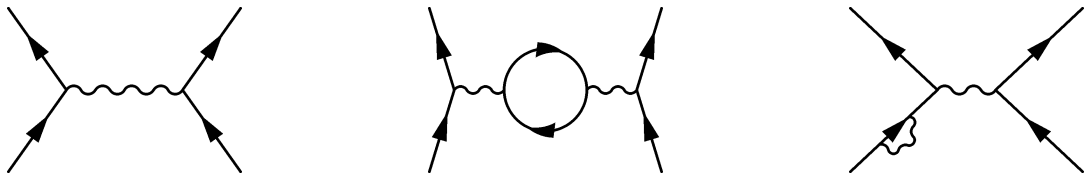

There is an additional difficulty with the Standard Model.  Most calculations in a quantum field theory are done perturbatively.  For example, the scattering cross section of two electrons at a given energy can be calculated up to a certain power of $\alpha$, the fine structure constant, which is the coupling constant for electromagnetism.  The calculation is represented pictorially with Feynman diagrams; the number of particle interaction vertices is related to the power of $\alpha$.  However, virtual particles and more complicated diagrams can also contribute to the process with higher powers of $\alpha$, several examples of which can be seen in Fig.~\ref{fig:4}.

\bigskip

As an example, the best anomalous magnetic moment of the electron calculation involves 891 diagrams.\cite{Kinoshita}  Thus most quantities have so-called ``quantum-loop" corrections (although some quantities, like the photon's mass, are protected by symmetries).  The fine-tuning problem arises when trying to calculate the mass of the Higgs particle; quantum loop corrections to the Higgs mass are quadratically divergent.  If one uses $10^{16}$ GeV as the scale at which the electroweak and strong forces combine to become a single unified force (which has been theorized but not seen), one requires an almost perfect cancellation on the order of 1 part in $10^{14}$ for the Higgs mass to come out at the electroweak scale ($\approx 150$ GeV).  This unnatural cancellation is a source of alarm for theorists and signifies that we lack an understanding of physics beyond the SM.

\section{Supersymmetry}

One possible extension to the Standard Model is supersymmetry (SUSY).  SUSY at its essence is an additional symmetry between fermions and bosons and can best be understood by beginning with the Coleman-Mandula theorem.  The Coleman-Mandula theorem states that the most general symmetries that a quantum field theory (QFT) can possess are Lorentz invariance (special relativity) and gauge symmetries like conservation of charge, lepton number, etc. (whose generators belong to Lie Algebras\cite{wessbagger}).  In other words, the Coleman-Mandula theorem is a ``no-go" theorem: a relativistic QFT can have no other symmetries.  In particular, there can be no change of the spin of particles.  That is, there is no way in the SM to change fermions to bosons or vice-versa.

However, in the mid-1970s two groups of physicists realized that the Coleman-Mandula theorem can be evaded.  Supersymmetry evades the restriction of the Coleman-Mandula theorem by generalizing and loosening the restriction on the types of symmetries of a QFT (in addition to Lie Algebras one can consider graded Lie Algebras whose operators anti-commute).  This additional symmetry allows for the inter-conversion of fermions and bosons.  Essentially, every fermion is now associated with a superpartner boson and every boson with a superpartner fermion; adding supersymmetry
to the standard model effectively doubles the number of particles.  Although doubling the number of particles may seem a hopeless complication, supersymmetry is very attractive theoretically for a number of reasons.

To begin with, supersymmetry may solve the hierarchy and the fine-tuning/naturalness problem in the Standard Model.  SUSY is new physics which acts at energies beyond the SM which helps to explain why the electroweak and Planck energy scales are so different. In terms of the fine-tuning problem, SUSY can explain why the Higgs mass and the Higgs vev are so small.   If SUSY were an exact symmetry of nature, then the mass of each SM bosonic particle must be equal to its superpartner
fermion mass. And since boson and fermion mass corrections in QFT calculations enter with opposite signs, they can cancel each other leading to a
``naturally" small Higgs mass and vev. Of course supersymmetry is a broken symmetry, meaning that the symmetry is no longer valid at the typical energies and background temperatures in the Universe today.  For example, we don't see a bosonic superpartner to the electron with~.511 MeV mass which would be a sign of unbroken supersymmetry.  Due to this breaking (which is not well understood) all superpartners must be extremely massive (much like the W and Z particles acquire mass in electroweak symmetry breaking while the photon remains massless).  In order to produce
acceptable corrections to the Higgs mass, the difference between boson and fermion masses must be of the order of 1 TeV.

Furthermore, precision measurements of Standard Model parameters at the LEP
collider show that using only the Standard Model particle content the strong, weak, and electromagentic forces do not seem to unify at energies of about $10^{16}$ GeV.  Particle physicists have long predicted that like the weak and electromagnetic forces which unify at energies of about $10^3$ GeV, the three quantum forces should merge to become a single Grand Unified Force.  However, if one adds the minimal particle content of supersymmetry, the couplings indeed seem to converge at a unification scale of $M \simeq 2 \times 10^{16}$ GeV.  \cite{unification,unification2,unification3}.  Additionally, supersymmetry is inherent in string theory, which currently is the only theory which has the possibility of unifying the quantum world with gravity.

And finally, and this is perhaps the most appealing characteristic of supersymmetry, the Standard Model with SUSY does in fact offer a viable dark matter candidate which we will discuss shortly.  The new particles generated by adding SUSY to the SM are shown below in Table~\ref{tab:2}. \\

\begin{table}[h]
\begin{center}

\begin{tabular}{l}
\textbf{Sfermions}
\end{tabular}

\begin{tabular}{|lcc|lcc|lcc|}
\hline ~Generation 1 &&& ~Generation 2 &&& ~Generation 3 &&\\
\hline ~Particle & Mass (GeV) & ~Charge~ & ~Particle & Mass (GeV) & ~Charge~ & ~Particle & Mass (GeV) & ~Charge~ \\
\hline ~up squark ($\tilde{u}$) & $> 379$ & +$\frac{2}{3}$ & ~charm squark ($\tilde{c}$) & $> 379$ & +$\frac{2}{3}$ & ~top squark ($\tilde{t}$) & $> 92.6$ & +$\frac{2}{3}$\\
~down squark ($\tilde{d}$) & $> 379$ & $-\frac{1}{3}$ & ~strange squark ($\tilde{s}$) & $> 379$ & $-\frac{1}{3}$ & ~bottom squark ($\tilde{b}$) & $> 89$ & $-\frac{1}{3}$\\
~selectron ($\tilde{e}$) & $> 73$ & $-$1 & ~smuon ($\tilde{\mu}$) & $> 94$ & $-$1 & ~stau ($\tilde{\tau}$) & $> 81.9$ & $-$1\\
~$e$ sneutrino ($\tilde{\nu}_e$) ~ & $> 95$ & 0 & ~$\mu$ sneutrino ($\tilde{\nu}_{\mu}$) & $> 94$ & 0 & ~$\tau$ sneutrino ($\tilde{\nu}_{\tau}$) & $> 94$ & 0\\
\hline
\end{tabular}

\begin{tabular}{l}
~ \\
\textbf{Gauginos}
\end{tabular}

\begin{tabular}{|lcl|}
\hline ~Particle & ~Mass (GeV)~ & Description\\
\hline ~Neutralinos ($\tilde{\chi}_{1-4}^0$) & $> 46$ & Mixture of photino ($\tilde{\gamma}$), zino ($\tilde{Z}$), and neutral higgsino ($\tilde{H}^0$)\\
       ~Charginos ($\tilde{\chi}_{1,2}^\pm$) & $> 94$ & Mixture of winos ($\tilde{W}^\pm$) and charged higgsinos ($\tilde{H}^\pm$)\\
       ~Gluinos ($\tilde{g}$) & $> 308$ & Superpartner of the gluon\\
\hline
\end{tabular}
\caption{The particles predicted by a supersymmetric extension of the Standard Model.  Limits on the masses of particles as last reported by the Particle Data Group.\cite{PDG}}
\label{tab:2}

\end{center}
\end{table}

When examining the particle content of the SM with SUSY, there are
several possible particles which could act as dark matter.  These
are the neutralino (a particle state which is a superposition of the neutral superpartners of
the Higgs and gauge bosons), the
sneutrino (the superpartner of the neutrino), and the gravitino (the
superpartner of the graviton which would come from a quantum theory of gravity).  All of these particles are electrically neutral and weakly interacting,
and thus are ideal WIMP-like candidates for dark matter.  However, sneutrinos
annihilate very rapidly in the early universe, and sneutrino relic densities are too low to be
cosmologically significant.\cite{sneutrino}  And gravitinos act as hot dark matter rather than cold dark matter, and  large scale structure observations are inconsistent with a
universe dominated by hot dark matter.\cite{gravitino}  This leaves the neutralino as a viable candidate.

But how can the neutralino, an extremely massive particle, exist today in sufficient numbers to make up the bulk of the dark matter (generically, massive particles decay into lighter ones)?  The answer lies in what is called R-parity.  In the Standard Model symmetries guarantee baryon and lepton number conservation; for this reason, the proton, the lightest baryon, cannot decay.   However, with the addition of supersymmetry, this is no longer generally true due to the presence of squarks and sleptons; recall, SUSY changes quarks and leptons into bosons and vice-versa, so baryon and lepton number are violated as a matter of course.  However, we know that the amount of baryon and lepton number violation (at least at low energies) must be extremely small due to sensitive tests.  An interesting property of SUSY is that if one writes down a theory without lepton and baryon number violating terms, no such terms will ever appear, even through quantum loop corrections (another advantage of SUSY is that certain types of quantities never get loop corrections).  Under this assumption, a new symmetry, called R-parity, may be conserved by a SUSY version of the SM.   We assign +1 R-parity for all Standard Model fields (including both Higgs fields), and -1 R-parity for all superpartners.  The
immediate consequence of R-parity conservation is that because
there are an even number of SUSY particles in every interaction, the lightest supersymmetric partner, the LSP, is stable and will not decay.  If this LSP is neutral, it is an excellent
candidate for dark matter.

In most SUSY versions of the standard model, the neutralino is the LSP and seems to be the most
promising dark matter candidate; the relic abundance of
neutralinos can be sizeable and of cosmological significance, and
detection rates are high enough to be accessible in the laboratory
but not high enough to be experimentally ruled out.  Thus the SM with SUSY offers a single
dark matter candidate: the neutralino.    Although at present not one supersymmetric particle has been detected in the laboratory, supersymmetry currently offers the best hope of modeling and understanding dark matter.  One clear advantage is that the minimal extension of the Standard Model using supersymmetry is well understood, and calculations, including dark matter densities and detection rates, can be performed.

\section{Exotic Candidates}

Although the neutralino and SUSY are well-motivated, other particle candidates for dark matter also exist.  The axion is a particle proposed in 1977 by Roberto Peccei and Helen Quinn to solve the so-called ``strong-CP problem."\cite{axion}  In a nutshell, the strong force Lagrangian contains a term that can give an arbitrarily large electric dipole moment to the neutron; since no electric dipole moment for the neutron has ever been observed, Peccei and Quinn postulated that a new symmetry prevents the appearance of such a term (much like a gauge symmetry keeps the photon massless).  They further theorized that this symmetry is slightly broken which leads to a new, very light scalar particle, the axion.  Although this particle is extremely light (theories place its mass in the $\mu$eV range), it can exist in sufficient numbers to act as cold dark matter.  Since axions should couple to photons, axions can be searched for with precisely tuned radio frequency (RF) cavities; inside the magnetic field of an RF cavity the axion can be converted into a photon which shows up as excess power in the cavity.  And in a unique blend of particle and astro-physics, limits on axions have been placed through observations of red giant stars; axions, if they existed, would offer another cooling mechanism which can be constrained by studying how quickly red giant stars cool.\cite{Raffelt} Although the axion has never been observed directly, several experiments such as ADMX and CARRACK are continuing the search and setting new limits on axion parameters.\cite{Duffy,Yamamoto}

If axions exist and SUSY is also correct, then the axino (the supersymmetric partner of the axion) is by a wide margin the LSP; neutralinos would decay into axinos through $\chi \rightarrow \tilde a + \gamma$.\cite{axinos} However, axinos would also act as hot dark matter and thus could not compose the bulk of the dark matter.

One final exotic particle candidate for dark matter comes from theories of extra spatial dimensions.  The idea that our universe could have extra spatial dimensions began in the 1920s with Theodor Kaluza and Oscar Klein; by writing down Einstein's general theory of relativity in five dimensions, they were able to recover four dimensional gravity as well as Maxwell's equations for a vector field (and an extra scalar particle that they didn't know what to do with).\cite{Kaluza,Klein}  Klein explained the non-observation of this extra fifth dimension by compactifying it on a circle with an extremely small radius (something like $10^{-35}$ cm so that it is completely non-observable).  Kaluza-Klein theories (as they came to be known) were an initial quest for a grand unified theory; however, the emergence of the weak and strong nuclear forces as fundamental forces of nature relegated this type of approach to unification to the drawing board.  However, in the late 1990s two new scenarios with extra dimensions appeared. Arkani-Hamed, Dvali, and Dimpoulos tried to solve the hierarchy problem by assuming the existence of large extra dimensions (initially on the scale of mm or smaller); they made the bold assertion that the electroweak scale is the only fundamental scale in nature and that the Planck scale appears so small due to the presence of these extra dimensions.\cite{Arkani}  Lisa Randall and Raman Sundum, on the other hand, proposed infinitely large extra dimensions that were unobservable at low energies; gravity, they explained, was weak precisely because it was the only force that could ``leak-out" into this extra dimension.\cite{Randall}

What do extra dimensions have to do with candidates for dark matter?  In theories in which extra dimensions are compactified, particles which can propagate in these extra dimensions have their momenta quantized as $p^2 \sim 1/R^2$, where $p$ is the particle's momentum and $R$ is the size of the extra dimension (we also use natural units were $\hbar=c=1$).  Therefore, for each particle free to move in these extra dimensions, a set of fourier modes, called Kaluza-Klein states, appears:

\begin{equation}
m^2 = \frac{n^2}{R^2} + m^2_0,
\end{equation}

\noindent where $R$ is the size of the compactified extra dimension, $m_0$ is the regular standard model mass of the particle, and $n$ is the mode number.  Each standard model particle is then associated with an infinite tower of excited Kaluza-Klein states.  If translational invariance along the fifth dimension is postulated, then a new discrete symmetry called Kaluza-Klein parity exists and the Lightest Kaluza-Klein particle (LKP) can actually be stable and act as dark matter; in most models the LKP is the first excitation of the photon.   Working within the framework of the Universal Extra Dimensions (UED) model \cite{UED}, Servant and Tait showed in 2003 that if the LKP has mass between 500 and 1200 GeV, then the LKP particle can exist in sufficient numbers to act as the dark matter\cite{Tait}.  Additional motivations for extra dimensional theories of dark matter include proton stability and the cancellation of gauge anomalies from three generations of fermions.  
\cite{HooperandProfumo}

Many other theories have been proposed to account for the universe's dark matter, most of which are not as promising as those already discussed. These include exotic candidates such as Q-balls, WIMPzillas, branons, and GIMPs among many others.\cite{Coleman,Cembranos, GIMPs} However, the neutralino remains the most studied and most theoretically motivated dark matter candidate.  In the next section we will discuss how particles like the neutralino (the SUSY LSP) can be produced in the early universe and how to determine if they exist today in sufficient density to act as the dark matter.

\section{Production in the Early Universe}

\subsection{Review of Big Bang Cosmology}

To understand how dark matter particles could have been created in the early universe, it is necessary to give a brief review of the standard model of cosmology.  The standard model of cosmology, or the ``Big Bang Theory" in more colloquial language, can be summed up in a single statement: the universe has expanded adiabatically from an initial hot and dense state and is isotropic and homogeneous on the largest scales.  Isotropic here means that the universe looks the same in every direction and homogeneous that the universe is roughly the same at every point in space; the two points form the so-called cosmological principle which is the cornerstone of modern cosmology and states that ``every comoving observer in the cosmic fluid has the same history" (our place in the universe is in no way special).  The mathematical basis of the model is Einstein's general relativity and it is experimentally supported by three key observations: the expansion of the universe as measured by Edwin Hubble in the 1920s, the cosmic microwave background, and Big Bang Nucleosynthesis.

The Friedmann equations are the result of applying general relativity to a four dimension universe which is homogeneous and isotropic:

\begin{eqnarray}
H^2 + \frac{k}{a^2} = \frac{8 \pi G}{3} \rho,  \\
\dot \rho = -3H \left(p + \rho \right),  \\
\frac{d}{dt} \left(s a^3 \right) = 0,
\end{eqnarray}

\noindent where $H$ is the Hubble constant (which gives the expansion rate of the Universe and is not really a constant but changes in time), $k$ is the spatial curvature of the universe today, $G$ is Newton's gravitation constant, and $p$ and $\rho$ are the pressure and energy density respectively of the matter and radiation present in the universe.  $a$ is called the scale-factor which is a function of time and gives the relative size of the universe ($a$ is defined to be 1 at present and 0 at the instant of the big bang), and $s$ is the entropy density of the universe.  The Friedmann equations are actually quite conceptually simple to understand.  The first says that the expansion rate of the universe depends on the matter and energy present in it.  The second is an expression of conservation of energy and the third is an expression of conservation of entropy per co-moving volume (a co-moving volume is a volume where expansions effects are removed; a non-evolving system would stay at constant number or entropy density in co-moving coordinates even through the number or entropy density is in fact decreasing due to the expansion of the universe).

The expansion rate of the universe as a function of time can be determined by specifying the matter or energy content through an equation of state (which relates energy density to pressure).   Using the equation of state $\rho = w p$, where $w$ is a constant one finds:

\begin{equation}
a(t) \propto \left(\frac{t}{t_0} \right)^{\frac{2}{3(1+w)}}
\end{equation}

\noindent where $t_0$ represents the present time such that $a(t_0)=1$ as stated earlier.  For non-relativistic matter where pressure is negligible, $w=0$ and thus $a \propto t^{2/3}$; and for radiation (and highly relativistic matter) $w=1/3$ and thus $a \propto t^{1/2}$.   Although the real universe is a mixture of non-relativistic matter and radiation, the scale factor follows the dominant contribution; up until roughly 47,000 years after the big bang, the universe was dominated by radiation and hence the scale factor grows like $t^{1/2}$.  Since heavy particles like dark matter were created before nucleosynthesis (which occurred minutes after the big bang), we shall treat the universe as radiation dominated when considering the production of dark matter.

\subsection{Thermodynamics in the Early Universe}

Particle reactions and production can be modeled in the early universe using the tools of thermodynamics and statistical mechanics.  In the early universe one of the most important quantities to calculate is the reaction rate per particle:

\begin{equation}
\Gamma = n \sigma v,
\label{equilib}
\end{equation}

\noindent where $n$ is the number density of particles, $\sigma$ is the cross section (the likelihood of interaction between the particles in question), and $v$ is the relative velocity.  As long as $\Gamma \gg H(t)$ we can apply equilibrium thermodynamics (basically this measures if particles interact frequently enough or is the expansion of the universe so fast that particles never encounter each other).
This allows us to describe a system macroscopically with various state variables: $V$ (volume), $T$ (temperature), $E$ (energy), $U$ (internal energy), $H$ (enthalpy), $S$ (entropy), etc. These variables are path independent; so long as two systems begin and end at the same value of a state variable, the change in that variable for both systems is the same.  The most relevant quantity in the early universe is temperature.  Since time and temperature are inversely correlated, i.e. the early universe is hotter (the exact relationship is that $T \propto 1/a$), we can re-write the Hubble constant and reaction rates and cross sections in terms of temperature.  The temperature will also tell us if enough thermal energy is available to create particles; for example, if the temperature of the universe is 10 GeV, sufficient thermal energy exists to create 500 MeV particles from pair production, but not 50 GeV particles.

Statistical thermodynamics can be used to derive relations for the energy density, number density, and entropy density of particles in the early universe in equilibrium.   To do so, Bose-Einstein statistics are used to describe distributions of bosons and Fermi-Dirac statistics are used to describe distributions of fermions. The main difference between the two arises from the Pauli exclusion principle which states that no two identical fermions can occupy the same quantum state at the same time. Bosons, on the other hand, \emph{can} occupy the same quantum state at the same time. Hence there are small differences in the quantities we compute for bosons and fermions.  To obtain the relevant statistical quantities, we begin with the distribution factor. The distribution factor $f_i(p)$, which gives the relative numbers of particles with various values of momentum $p$ for a particle species $i$, is given by

\begin{equation}
f_i(p) = \frac{1}{\textrm{e}^{(E_i-\mu_i)/T} \pm 1},
\end{equation}

\noindent where $E_i=\sqrt{m_i^2+p^2}$, $\mu_i$ is the chemical potential of species $i$ (energy associated with change in particle number), and the +1 case describes bosons and the -1 case fermions. (You might notice that the Boltzmann constant $k$ is missing from the denominator of the exponential. To simplify things, cosmologists use a system of units where $\hbar=k=c=1$.) The distribution factor can be used to determine ratios and fractions of particles at different energies, as well as the number and energy densities which are given by the integrals

\begin{equation}
n_i = \frac{g_i}{(2 \pi)^3} \int f_i(p) \textrm{d}^3p = \frac{g_i}{2 \pi^2} \int p^2 f_i(p) \textrm{d}p,
\end{equation}

\noindent and

\begin{equation}
\rho_i = \frac{g_i}{(2 \pi)^3} \int E_i f_i(p) \textrm{d}^3p = \frac{g_i}{2 \pi^2} \int p^2 E_i f_i(p) \textrm{d}p,
\end{equation}

\noindent where $\textrm{d}^3p = 4 \pi p^2 \textrm{d}p$ and $g_i$ is the number of degrees of freedom. The degrees of freedom are also called the statistical weights and basically account for the number of possible combinations of states of a particle. For example, consider quarks. Quarks have two possible spin states and three possible color states, and there are two quarks per generation. So, the total degrees of freedom for the quarks in the Standard Model  are $g_q=2\times3\times2=12$ with an addition $g_{\bar q} = 12$ for the anti-quarks. Each fundamental type of particle (and associated anti-particle) has its own number of degrees of freedom, which enter into the calculations of number and energy densities. The known particles of the SM (plus the predicted Higgs boson) have a total of 118 degrees of freedom.

The integrals for number and energy densities can be solved explicitly in two limits: 1) the relativistic limit where $m \ll T$ and 2) the non-relativistic limit where $m \gg T$.  For nonrelativistic particles where $m \gg T$,

\begin{equation}
n_{NR} = g_i \left(\frac{mT}{2\pi}\right)^{3/2} \textrm{e}^{-m/T},
\end{equation}

\noindent which is a Maxwell-Boltzmann distribution (no difference between fermions and bosons) and

\begin{equation}
\rho_{NR} = m_n.
\end{equation}

\noindent For relativistic particles, on the other hand, where $m \ll T$,

\begin{equation}
n_R=\left\{
\begin{array}{l l}
    \frac{\zeta(3)}{\pi^2}g_iT^3 ~~~~\textrm{for bosons}\\
    \frac{3}{4} \frac{\zeta(3)}{\pi^2}g_iT^3 ~~~~\textrm{for fermions}
\end{array}\right.
\label{bosons}
\end{equation}

\noindent and

\begin{equation}
\rho_R=\left\{
\begin{array}{ll}
    \frac{\pi^2}{30}g_iT^4 ~~~~\textrm{for bosons}\\
    \frac {7}{8} \frac{\pi^2}{30}g_iT^4 ~~~~\textrm{for fermions}
\end{array}\right.
\end{equation}

\noindent where $\zeta$ is the Riemann Zeta function.  These results show that at any given time (or temperature) only relativistic particles contribute significantly to the total number and energy density; the number density of non-relativistic species are exponentially suppressed.  As an example, knowing that the CMB photons have a temperature today of 2.73 K, we can use Eq.~(\ref{bosons}) to calculate $n_\gamma \approx 410$ photons/cm$^3$.

\subsection{Particle Production and Relic Density: The Boltzmann Equation}
\begin{figure}[htp]
\includegraphics[width=0.50\textwidth]{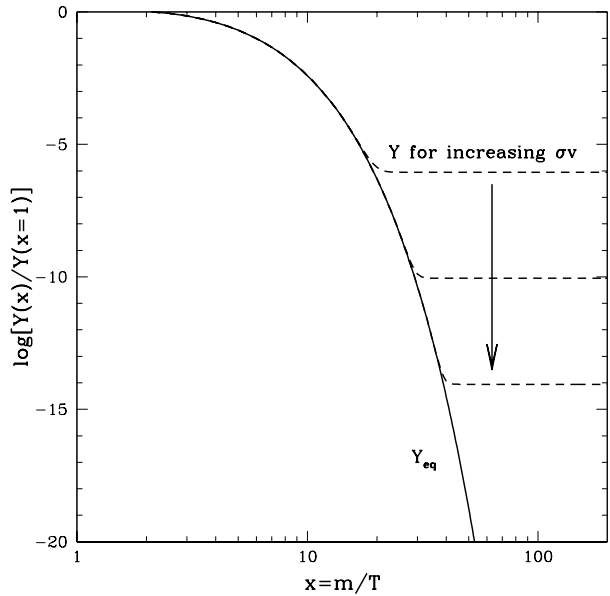}
\caption{The evolution of $Y(x)/Y(x=1)$ versus $x=m/T$ where $Y=n/s$.} \label{fig:2}
\end{figure}

In the early universe, very energetic and massive particles were created and existed in thermal equilibrium (through mechanisms like pair production or collisions/interactions of other particles).  In other words, the processes which converted heavy particles into lighter ones and vice versa occurred at the same rate.  As the universe expanded and cooled, however, two things occurred: 1) lighter particles no longer had sufficient kinetic energy (thermal energy) to produce heavier particles through interactions and 2) the universe's expansion diluted the number of particles such that interactions did not occur as frequently or at all.   At some point, the density of heavier particles or a particular particle species became too low to support frequent interactions and conditions for thermal equilibrium were violated; particles are said to ``freeze-out" and their comoving number density (no longer affected by interactions) remains constant.  The exact moment or temperature of freeze-out can be calculated by equating the reaction rate, Eq.~(\ref{equilib}), with the Hubble (expansion) rate.   The density of a specific particle at the time of freeze-out is known as the relic density for this particle since its abundance remains constant (see Fig.~\ref{fig:2}).

We know that a supersymmetric particle like the neutralino is a good dark matter candidate: it is electrically neutral, weakly interacting, and massive.  It is also produced copiously in the early universe.  But can a particle like the neutralino (or a generic WIMP) still exist in the present day universe with sufficient density to act as dark matter?  To answer this question we will explore how precisely to compute relic densities.

The  Boltzmann equation gives an expression for the changing number density of a certain particle over time, d$n/$d$t$. To formulate the Boltzmann equation for a particle species, relevant processes must be understood and included. For supersymmetric particles, like the neutralino, there are four: 1) the expansion of the universe, 2) coannihilation, in which two SUSY particles annihilate with each other to create standard model particles, 3) decay of the particle in question, and 4) scattering off of the thermal background. Each of these processes then corresponds respectively to a term in the Boltzmann equation for d$n_i/$d$t$ where there are $N$ SUSY particles:

\begin{eqnarray}
\label{Boltzmann1}
\frac{\textrm{d}n_i}{\textrm{d}t} = - 3Hn_i - \displaystyle \sum_{j=1}^N \langle \sigma_{ij}v_{ij} \rangle (n_in_j-n_i^{eq}n_j^{eq}) - \displaystyle \sum_{j \ne i} \left[ \Gamma_{ij}(n_i-n_i^{eq})-\Gamma_{ji}(n_j-n_j^{eq})\right] \nonumber
\\
-\displaystyle \sum_{j \ne i} \left[ \langle \sigma^{\prime}_{Xij} v_{ij} \rangle (n_in_X-n_i^{eq}n_X^{eq}) - \langle \sigma^{\prime}_{Xji} v_{ji} \rangle (n_jn_X-n_j^{eq}n_X^{eq}) \right].
\end{eqnarray}

\noindent In the case of supersymmetry (which we will focus on for the rest of this section) the Boltzmann equation can be simplified by considering R-parity.  If we also assume that the decay rate of SUSY particles is much faster than the age of the universe, then all SUSY particles present at the beginning of the universe have since decayed into neutralinos, the LSP. We can thus say that the abundance of neutralinos is the sum of the density of all SUSY particles ($n = n_i+...+n_N$). Note, then, that when we take the sum in Eq.~(\ref{Boltzmann1}), the third and fourth terms cancel. This makes sense, because conversions and decays of SUSY particles do not affect the overall abundance of SUSY particles (and therefore neutralinos), and thus make no contribution to d$n/$d$t$. We are left with

\begin{equation}
\label{Boltzmann2}
\frac{\textrm{d}n}{\textrm{d}t} = - 3Hn - \langle \sigma_{eff} v \rangle (n^2 - n_{eq}^2),
\end{equation}

\noindent where $n$ is the number density of neutralinos, $\langle \sigma_{eff} v \rangle$ is the thermal average of the effective annihilation cross section $\sigma_{eff}$ times the relative velocity $v$, and $n_{eq}$ is the equilibrium number density of neutralinos. A quick note should be made about the thermal average $\langle \sigma_{eff} v \rangle$ and the added difficulty behind it. As thoroughly described by M. Schelke, coannihilations between the neutralinos and heavier SUSY particles can cause a change in the neutralino relic density of more than 1000\%, and thus should not be left out of the calculation.\cite{Schelke} The thermal average is then not just the cross section times relative velocity of neutralino-neutralino annihilation, but of neutralino-SUSY particle annihilations as well; many different possible reactions must be considered based upon the mass differences between the neutralino and other SUSY particles.

By putting Eq.~(\ref{Boltzmann2}) in terms of $Y=n/s$ and $x=m_{\chi}/T$ where $T$ is the temperature to simplify the calculations, we obtain the form

\begin{equation}
\label{Boltzmann3}
\frac{\textrm{d}Y}{\textrm{d}x} = - \sqrt{\frac{\pi}{45 G}} \frac{g_*^{1/2} m_{\chi}}{x^2} \langle \sigma_{eff} v \rangle (Y^2 - Y_{eq}^2),
\end{equation}

\noindent where $g_*^{1/2}$ is a parameter which depends on the effective degrees of freedom. Eq.~ (\ref{Boltzmann3}) can then be integrated from $x=0$ to $x_0=m_{\chi}/T_0$ to find $Y_0$ (which will be needed in Eq.~(\ref{relicdens}) for the relic density).\cite{Gondolo}

Fig.~(\ref{fig:2}) (adapted from Kolb and Turner's excellent treatment of this topic\cite{kolbturner})
plots an analytical approximation to the Boltzmann Equation and illustrates several key points.  The y-axis is essentially (or at least proportional to) the relic density; the solid line is the equilibrium value and the dashed-line is the actual abundance.  Notice that at freeze-out, the actual abundance leaves the equilibrium value and remains essentially constant; the equilbrium value, on the other hand, continues to decrease so freeze-out is key to preserving high relic densities.  Furthermore, the larger the annihilation cross section, $\langle \sigma_a v \rangle$, the lower the relic density; this makes sense since the more readily a particle annihilates, the less likely it will exist in large numbers today.  This has been paraphrased by many as ``the weakest wins" meaning that particles with the smallest annihilation cross sections will have the largest relic densities.

Using the value of $Y_0$ described above, the relic density of neutralinos is given by

\begin{equation}
\label{relicdens}
\Omega_{\chi} h^2 = \frac{h^2 m_{\chi} s_0 Y_0}{\rho_{crit}} = 2.755 \times 10^8 \frac{m_{\chi}}{\textrm{GeV}} Y_0,
\end{equation}

\noindent where $m_{\chi}$ is the mass of the neutralino, $s_0$ is the entropy density today (which is dominated by the CMB photons since there are about $10^{10}$ photons per baryon in the universe), and $Y_0$ is the result of integrating the modified version of the Boltzmann equation.  For a more thorough discussion of the Boltzmann equation and the relic density of neutralinos, consult M. Schelke and J. Edsj\"{o} and P. Gondolo.\cite{Schelke,Edsjo}

Recall that the difference between the matter and baryon densities (as determined by WMAP) was $\Omega_{\chi} h^2 =0.11425 \pm 0.00311$.  Can particle models of dark matter like SUSY or Kaluza-Klein theories produce dark matter in sufficient quantities to act as the bulk of the dark matter?  The answer generically is yes.  Although SUSY theories cannot predict a single dark matter relic density due to the inherent uncertainty in the input parameters, the Standard Model plus supersymmetry does produce a wide range of models some of which have the expected dark matter density.   For example, MSSM models (the minimal supersymmetric standard model) yield dark matter relic densities from 10$^{-6}$ of the WMAP results to some which overclose the universe; neutralino masses for the models which give a correct relic abundance typically lie in the range $50-1000$ GeV.   A similar calculation can be performed for Kaluza-Klein dark matter in the Universal Extra Dimension scenario; Hooper and Profumo report that with the first excitation of the photon as the LKP a relic abundance can be obtained in the proper range of $0.095 < \Omega_{dm} < 0.129$ for LKP masses between 850 and 900 GeV.\cite{UED}

To summarize, statistical thermodynamics allows us to model conditions in the early universe to predict relic abundances of dark matter particles like neutralinos.  What is remarkable is that such models produce dark matter abundances consistent with the range provided by cosmological measurements.  Theories like SUSY and Kaluza-Klein theories are so powerful, in particular, because such calculations (even given a wide range of uncertainty in several parameters) are possible.  Armed with the required relic density of dark matter and the various models proposed by theorists, experiments can search for dark matter in specific areas of allowed parameter space. In the next section we will discuss the methods and progress of the various detection schemes and experiments.

\section{Detection Schemes}
Detecting (or creating) dark matter is key in determining its
properties and the role of dark matter in the formation of structure in the universe.  Many
experiments have searched and are currently searching for
a signal of WIMP-like dark matter (many specifically for neutralinos) and each
uses a different detection method.  Although producing dark matter in
a particle accelerator would be ideal (we would have better control and the experiment
would be repeatable), but other methods to find
dark matter, coined direct and indirect detection, also continue to be important
in the search.  Due to the summary nature of this article, only a brief
overview of this large topic will be presented here.  For a
more thorough review of detection techniques and results,
refer to L. Baudis and H. V. Klapdor-Kleingrothaus for direct
detection and J. Carr \emph{et al.} for indirect.\cite{Baudis,Carr}  In this discussion we have assumed that the local dark matter distribution is at rest in the Milky Way halo.  However, other possibilities, including dark mater clumps and velocity streams due to incomplete virialization are possible.\cite{Diemand}

\subsection{Production in Accelerators}

Producing and detecting dark matter particles in an accelerator
would be a huge step toward confirming the  existence of
dark matter (though it would not, intriguingly, verify that the produced particle acts as the vast amount of dark matter in the universe detected through astrophysical means; in principle, dark matter
in the universe need not be comprised of a single component).  
If we assume that R-parity is conserved and that the dark matter is the neutralino and thus the LSP, then a signal in an accelerator will have several distinctive features. When SUSY
particles are created, they will decay to the LSP and most likely
escape the detector (similar to a neutrino - remember SUSY particles
interact very weakly with regular matter). As the LSP leaves the collision space, it will
carry with it energy and momentum which can be detected as missing
energy and momentum.  Similar signatures of missing energy would be detected if the dark matter were
Kaluza-Klein excitations or other exotic particles.  

Although direct evidence for SUSY or other exotic particles hasn't been seen yet,
there are certain processes which depend heavily on whether they
exist. For example, the radiative quark decay process ($s\rightarrow
b\gamma$) and the anomalous magnetic moment of the muon constrain
the possible masses of SUSY particles. The constraints obtained from
these and other experiments, however, are highly model dependent so it
is therefore difficult to make any general claims about them.  Since there is such uncertainty in the theory, we will generically refer in the following sections to dark matter particles as WIMPs (Weakly Interacting Massive Particles).

\subsection{Direct Detection}

The basic idea of direct detection is simple: set up a very
sensitive device, containing a large amount of some element,
which can detect very small motions and interactions of the
atoms within it.  If dark matter is everywhere in the universe,
then it should be traveling around (and through) the earth,
and therefore a detection apparatus, at all times.  
Although dark matter is weakly interacting, it may occasionally bump into
the nucleus of a detector atom and deposit some energy which
can be sensed by the detector.  To get an idea of how much
energy a WIMP would deposit, we first estimate that WIMPs are
moving at velocities of about 220 km/s and their masses are
somewhere around 100 GeV.  We then crudely find a WIMP's kinetic
energy using

\begin{equation}
T = \frac{1}{2}mv^2 = \frac{1}{2}(1.783 \times 10^{-25}~\mbox{kg})(220,000~\mbox{m/s})^2 \approx 4.314 \times 10^{-15}~\mbox{J} \approx 26.9~\mbox{keV}.
\end{equation}

\noindent This is the upper limit of energy that a 100 GeV
neutralino traveling at 220 km/s could deposit in the
detector; the actual amount would almost certainly be
smaller, since it is unlikely for a weakly-interacting
particle to be completely stopped within the detector.
Natural radioactivity generally emits MeV energies, making
a keV increase in energy due to nuclear scattering nearly
impossible to find.  For this reason, direct detection devices
must be radioactively clean and shielded from particles
that may make detection of WIMPs difficult.

The recoil energy $E$ of a WIMP with mass $m$ scattering off
of a nucleus of mass $M$ can be more precisely found with the
expression

\begin{equation}
E = \frac{\mu^2 v^2}{M}(1-\mbox{cos} \theta),
\end{equation}

\noindent where $\mu = mM/(m+M)$ is the reduced mass, $v$
is the speed of the WIMP relative to the nucleus, and $\theta$
is the scattering angle.

A WIMP signal should have specific characteristics. First,
events should be uniformly distributed throughout
the detector given that the local dark matter density is
thought to be fairly homogeneous and the cross section of
interaction remains constant.  Secondly, a WIMP-nucleus interaction
should be a single-site event, whereas an event from cosmic rays or naturally occuring
radioactivity can be multi-site.  For this reason, detectors have an
``anti-coincidence veto system" which makes sure events
that occur extremely close together (within nanoseconds),
suggesting that they are caused by the same incoming particle,
are not counted as caused by other WIMPs.  Detection rates
should also vary at different times of the year due to the
earth moving with or against the velocity of dark matter in
the galaxy.  This depends on the Earth's velocity, given by

\begin{equation}
v_E = 220 \mbox{ km/s } \Big(1.05 + 0.07 \cos (\omega(t-t_0))\Big),
\end{equation}

\noindent where $\omega$ is $2\pi$/year and $t_0$ is June 2.
As a result, the variation of WIMP flux over a year is only
about 7\%, meaning that many events would be required to see
such a small modulation.  These among other indications help
detection experiments decide whether received signals really
are WIMPs or not.

The interaction of a WIMP with the detector material can be
classified by two characteristics: elastic or inelastic, and
spin-dependent or spin-independent.

\begin{itemize}

\item Elastic and inelastic scattering: Elastic scattering is
the interaction of a WIMP with the nucleus as a whole.  This
causes the nucleus to recoil and, as we have seen, would deposit
energies of around 25 keV.  In inelastic scattering all of the energy does
not go into nuclear recoil; instead the nucleus is excited to a higher
energy state (for example, the 5/2$^+$ state in $^{73}$Ge) which then decays
by photon emission.  If the excited state is long-lived enough, the decay signal can be separated from the nuclear recoil event; this leads to better background discrimination.  However, inelastic scattering
cross sections are generally smaller than elastic scattering cross sections due to a lack
of coherence (the interaction is with individual nucleons rather than with the nucleus as a whole).\cite{EngelandVogel}

\item Spin-dependent and spin-independent scattering: Spin-dependent
(``axial-vector") scattering results from the coupling of a
WIMP's spin with the spin content of a nucleon.  Spin-independent
(``scalar") does not depend on this and has the advantage of
higher cross sections with larger nuclei (because of coherence where the WIMP interacts
with the nucleus as a whole).

\end{itemize}

\noindent A recoil event can then be further categorized,
taking on one of three forms:

\begin{itemize}
\item Phonon/Thermal: A vibration (detected as a rise in
temperature) in the crystal lattice of the detector, caused
by the slight movement of a nucleus off which a WIMP has
recoiled.  An extremely sensitive thermometer system is
located around the detector, allowing any temperature
variation to be recorded.

\item Ionization: An incident particle gives an electron
in the detector enough energy to escape the pull of its
nucleus.  A small electric field is set up in the detector
to ``push" the new charge to a detector wall where it can be
registered and counted as an ionization event.

\item Scintillation: Caused when an electron absorbs enough
energy to climb to a higher energy state.  After a short time, the electron will
lose this energy by emitting a photon, which is then
gathered by photomultipliers and converted to an electric
signal so it can be analyzed.
\end{itemize}

\noindent A detector is generally set up to sense two of
these WIMP signals.  By doing so, background events can
be recognized on an event-by-event basis and discarded,
allowing possible dark matter signatures to be counted and
analyzed.

To calculate the number of recoil events $N$ expected in
a detector within a range of recoil energy ($E_1, E_2$),
we take a sum over the nuclear species $i$ in the detector:

\begin{equation}
N_{E_1-E_2} = \sum_i \int_{E_1}^{E_2} \frac{\mbox{d}R_i}{\mbox{d}E} \varepsilon_i(E)~\mbox{d}E,
\end{equation}

\noindent where d$R_i/$d$E$ is the expected recoil rate
per unit mass of $i$ per unit nucleus recoil energy and
per unit time, and $\varepsilon_i(E)$ is the effective
exposure of $i$ in the detector.  d$R_i/$d$E$ is given by

\begin{equation}
\frac{\mbox{d}R_i}{\mbox{d}E}=\frac{\rho\sigma_i|F_i(E)|^2}{2m\mu_i^2} \int_{v>\sqrt{M_i E / 2\mu_i^2}} \frac{f(\vec{v},t)}{v}~\mbox{d}^3v,
\end{equation}

\noindent where $\rho$ is the local halo dark matter
density, $\sigma_i$ is the WIMP-nucleus cross section, $F_i(E)$ is the nuclear form factor
which takes into account that a nucleus is not a simple
point particle, $m$ is the WIMP mass, $\mu_i$ is the
reduced mass, $v$ is the velocity of the WIMP with respect
to the detector, $M_i$ is the mass of a nucleus of species
$i$, $E$ is the recoil energy, and $f(\vec{v},t)$ is the
WIMP velocity distribution (generally assumed to be a
Maxwell-Boltzmann distribution) in the reference frame
of the detector.  The nuclear physics uncertainties are locked into $F_i(E)$ while the
astrophysical uncertainties lie in the WIMP velocity distribution.  $\mathcal{E}_i(E)$ is given by

\begin{equation}
\mathcal{E}_i(E) = \mathcal{M}_i T_i \epsilon_i(E),
\end{equation}

\noindent where $\mathcal{M}$ is the total mass of nuclei
of species $i$ in the detector that has been active for a
time $T_i$, and $\epsilon_i(E)$ is the counting efficiency
for nuclear recoils of energy $E$.

We can see from these expressions that a detector should
ideally have a large mass with which to receive signals,
be operational for a long period of time, and be properly
shielded against background radiation.  An upper limit can
be put on the WIMP-nucleus cross section by comparing
expected events (using the above expressions) to observation.
Any negative results from direct detection experiments are not
wasted time and effort; instead, we can say that the
WIMP does not exist in a certain tested area of the
parameter space (for a given dark matter theory) and look toward more sensitive areas.
Fortunately, experiments are reaching more advanced detection
techniques and are approaching the parameter space in which
WIMPs are believed to exist.

While many others are in operation worldwide, the three direct
detection experiments which have yielded the best (most constrained)
results for the spin-independent WIMP-nucleus cross section
are the Cryogenic Dark Matter Search (CDMS II) in the Soudan
Mine, the UK Dark Matter Collaboration's ZEPLIN-I (ZonEd
Proportional scintillation in LIquid Noble gases) in the
Boulby Mine, and XENON10 at the Gran Sasso Underground Laboratory.

\begin{center}
\begin{figure}[htp]
\includegraphics[width=0.50\textwidth]{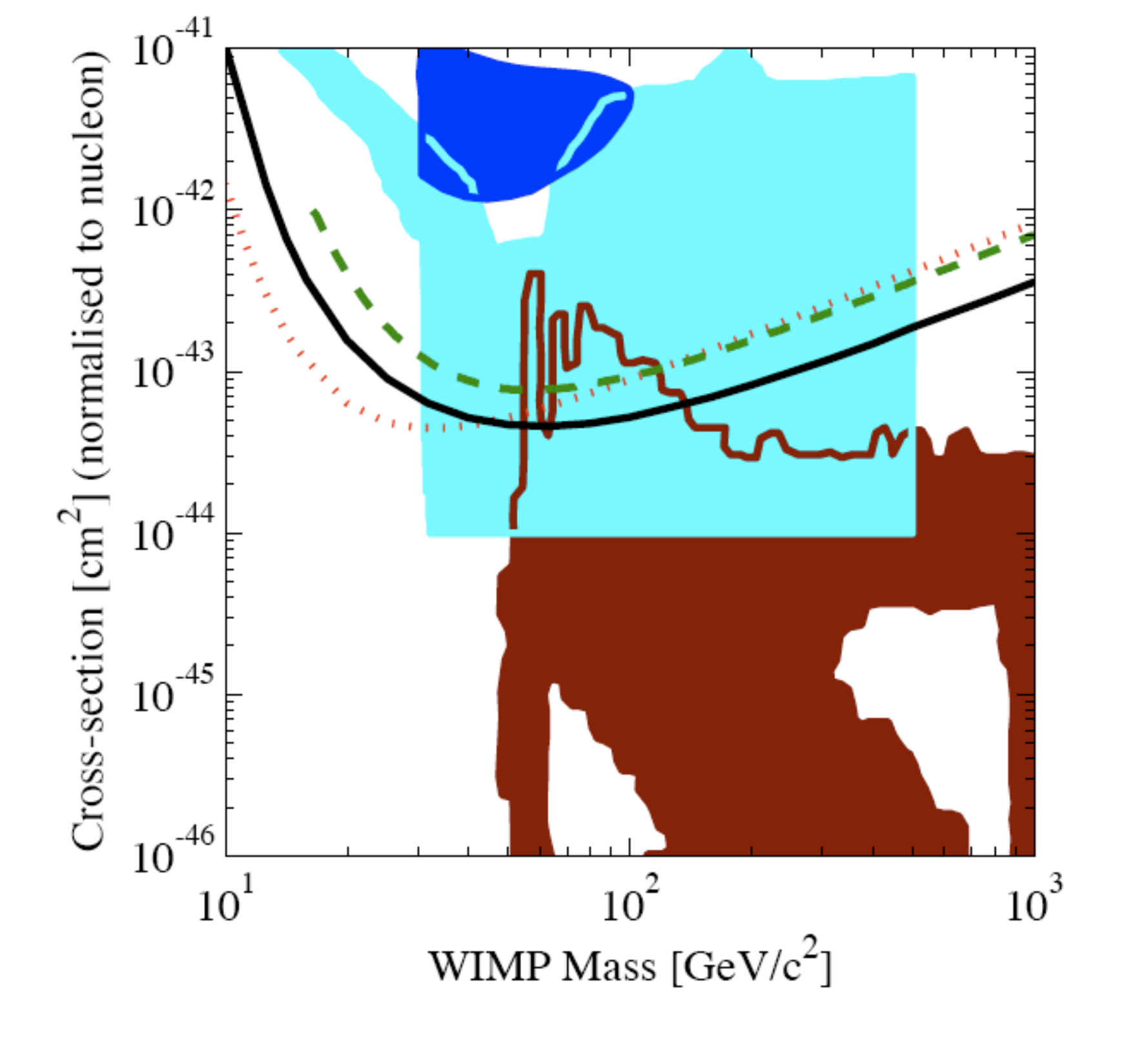}
\caption{Spin Independent cross section versus WIMP mass with exclusion curves.  The blue(top shaded) area is the DAMA claimed discovery region.  The remaining shaded areas are regions of parameter space in which theory predicts dark matter candidates.  The exclusion curves are: dashed - ZEPLIN III, solid line - CDMS II, dotted line - XENON 10.  Models above the exclusion curves are experimentally ruled out.  This plot was produced with the Dark Matter Limit Plot Generator.\cite{plotter}} \label{exclusion_curves}
\end{figure}
\end{center}

Each of these three collaborations uses a different method to look
for WIMPs. CDMS II uses a set of 250 gram Ge detectors
and 100 gram Si detectors cooled to less than 50 mK.
Each apparatus is classified as a ZIP (Z-dependent Ionization
and Phonon) detector.  While WIMPs recoil off of nuclei,
background particles scatter off of electrons; the ZIP detectors
are able to discriminate between the two events.\cite{CDMS2}  ZEPLIN-III uses
the scintillation properties of liquid Xe to detect WIMPs,
shielded by lead and a liquid scintillator veto to reduce background
radiation.\cite{ZEPLIN}  XENON10 also uses liquid Xe to detect scintillation
and ionization events.\cite{XENON} The exclusion curves from these
collaborations are shown in Fig.~(\ref{exclusion_curves}).  In Fig.~(\ref{exclusion_curves}), the nature and extent of the shaded regions from theory depend heavily on the individual supersymmetric theories; we mean only to show here that there are theories which predict WIMP candidates that are within the reach of current and planned direct detection experiments.

One experiment, the DAMA collaboration, detected an annual
modulation in scattering events that is around the expected
7\%.\cite{DAMA}  This is puzzling, however, because no other
direct detection experiment has found such a signal.  P.
Gondolo and G. Gelmini have given possible reasons for this;
for example, the WIMP velocity may be larger than the DAMA
thresholds, but smaller than the thresholds of other
detectors.\cite{Gondolo}  However, DAMA's conclusions still
remain controversial.

Other collaborations focusing on spin-independent direct
detection include CRESST and EDELWEISS.\cite{CRESST,EDELWEISS}
Many other future projects have been proposed, such as GENIUS
and SuperCDMS, which is planned to be able to probe nearly all
split-supersymmetry model parameter space.\cite{GENIUS,SuperCDMS}

Although spin-independent scattering has larger interaction
rates in most SUSY models, spin-dependent scattering can explore
the parameter space where scalar interaction is less probable.
For this reason, experiments searching for spin-dependent
interactions have been able to set competitive upper limits on
WIMP interactions. For example, $^{73}$Ge makes up 7.73\% of
the CDMS Collaboration's natural germanium detectors, and $^{29}$Si
accounts for 4.68\% of the natural silicon, both which have
nonzero spin.  As a result, CMDS's spin-dependent experiments
have placed the best current upper limit on WIMP-neutron
interactions.\cite{CDMS3}  Other direct detection experiments
employing spin-dependent techniques include PICASSO and
NAIAD.\cite{PICASSO,NAIAD}

\subsection{Indirect Detection}
In supersymmetry, for example, neutralinos are classified as Majorana particles (they are
their own antiparticle) and therefore annihilate with each
other, giving off various products which we can detect.  Hence a potential signal for the
existence of dark matter is WIMP-WIMP annihilation.  This
technique is called ``indirect detection" since we are not
actually detecting the WIMPs themselves.  Because the annihilation
rate of WIMPs is proportional to the square of the dark
matter density ($\Gamma_A \propto \rho_{DM}^2$), natural places
to look for dark matter annihilations are those expected to
have high WIMP densities, such as the sun, earth, and galactic
center.  Annihilation products include gamma-rays, neutrinos,
and antimatter.

\subsubsection{Gamma-rays}
Gamma-rays from WIMP annihilation are believed to occur
most frequently in the galactic center.  One way this process
can take place is through a WIMP annihilation yielding
a quark and anti-quark, which then produce a particle jet
from which a spectrum of gamma-rays is released.  The quark anti-quark
fragmentation process has been thoroughly studied at accelerators and is well understood;
the creation and propagation of gamma rays from such a jet is a fairly predictable process (compared to, for example, the ``random walk" of charged antimatter particles
through space).  A second form of gamma-ray production is
the decay of WIMPs directly to gamma-rays, $\chi \chi \rightarrow 
\gamma \gamma \mbox{ or } \gamma Z$), which produces gamma-rays
(a ``gamma-ray line") that are proportional to the mass of the
WIMPs involved.  Since typical WIMP masses can be on the
order of 100s of GeV, these are extremely high energy gamma rays.
Although the flux is small and quite difficult
to detect, observing such a gamma-ray line would be an obvious
indication for dark matter annihilation and the WIMP mass (often
referred to as the ``smoking gun").

As a gamma-ray enters an indirect detection device, it first
passes through an anti-coincidence shield which limits the
amount of charged particles entering the detector.  The
gamma-ray then encounters ``conversion foils," which are thin
sheets of heavy nuclei that convert the photon to a $e^+e^-$
pair.  A calorimeter tracks the energies of the positron and
electron, while particle tracking detectors measure their
trajectories.  The signature of a gamma-ray event is then
a registered energy with nothing triggering the anti-coincidence
shield, and the indication of two particles (the positron and
electron) coming from the same location.

The EGRET Collaboration reported an excess of gamma-rays
in 1998, pointing toward already accepted characteristics
of dark matter: a 50-70 GeV WIMP mass and a ring of concentrated
dark matter at a radius of 14 kpc from the galactic center
(which would nicely answer for our flat rotation curves).\cite{EGRET}
This discovery is initially encouraging, but as Bergstr\"{o}m
\emph{et al.} have shown, observed antiproton fluxes would have
to be much larger if these excess gamma rays are being produced
by neutralino or generic WIMP self annihilation.\cite{Bergstrom}  For this
reason and others, EGRET's results remain controversial.

\subsubsection{Neutrinos}
Neutrinos can be another important product of WIMP
annihilation.  As WIMPs travel through the universe and
through matter, they lose small amounts of energy due to
scattering off of nuclei.  Therefore, WIMPs can gather at
the centers of large gravitating bodies, increasing their
density until their annihilation rate equals half the
capture rate (two WIMPs are needed for annihilation, where
only one is needed for capture).  For many of the primary
particle physics models, the WIMP annihilation and capture
rates are at (or nearly at) equilibrium in the sun, a
conveniently ``close" object to observe.  This equilibrium
should allow for a steady annihilation rate, and therefore
a constant flow of neutrinos emanating from within the sun (we study only the neutrinos
and not other products of annihilation because neutrinos interact so weakly that most
escape from the sun or body in question).
Why not, then, study neutrinos coming from WIMP
annihilations within the earth, an even closer gravitating
body?  The earth (in most models) has not reached such an equilibrium and
thus does not provide a flux of neutrinos; it is less
massive than the sun, so it causes less WIMP scattering
and a much smaller gravitational potential well.  Neutrino
telescopes therefore usually focus on neutrino flux coming
from the sun, rather than the earth.

The differential neutrino flux from WIMP annihilation is given by

\begin{equation}
\frac{dN_\nu}{dE_\nu} = \frac{\Gamma_A}{4 \pi D^2} \sum_f B^f_X \frac{dN^f_\nu}{dE_\nu}
\end{equation}

\noindent where $\Gamma_A$ is the annihilation rate of WIMPs in the Sun and/or Earth, D is the distance of the detector from the source (the central region of the Sun and/or Earth), $f$ is the WIMP pair annihilation final states, and $B^f_X$ are the branching ratios of the final states.  $dN^f_\nu/dE_\nu$ are the energy distributions of neutrinos generated by the final state $f$.\cite{DarkSUSY_full}  
Depending on the WIMP's mass and composition, annihilation
processes include $\chi \chi \rightarrow t \bar{t},~b
\bar{b},~c \bar{c},~ZZ,~W^+ W^-,~\mbox{and}~\tau^+ \tau^-$,
which then decay to neutrinos among other products.  For
neutralinos or generic WIMPs lighter than $W^{\pm}$, annihilation to $b\bar{b}$
and $\tau^+\tau^-$ are the most common processes, yielding
neutrinos with energies around 30 GeV.  WIMPs with
higher masses annihilate to Higgs and gauge bosons, top and
bottom quarks, and muons, leading to neutrinos of masses
that are much easier to detect (about half of the WIMP mass).
Detection, then, depends heavily on the WIMP mass, as well
as the annihilation rate, density within the sun, and other
factors.

As neutrinos pass through the earth, they sometimes interact
with the hydrogen and oxygen and other atoms around the optical modules
of a neutrino detector.  Electrons, muons, and taus produced by such events are extremely energetic
and are traveling faster than the speed of light in the medium; the particles are then detected optically due to the Cherenkov radiation they emit.    

Because neutrinos are so weakly interacting, neutrino telescopes must be massive to
detect a significant signal.  AMANDA-II is a
neutrino detector 1500 to 2000 meters underground within the
ice of the South Pole where Cherenkov radiation can travel
and be seen easily by optical modules.  This experiment has
not detected statistically significant results from the
direction of the sun, but has placed helped place firm limits
on the muon flux.\cite{AMANDA}  A future experiment (expected
to be fully completed in 2011), IceCube, will integrate
AMANDA into a much larger detection experiment, with 7200
optical modules and a detector volume of a cubic
kilometer.\cite{IceCube}  Super-Kamiokande (``Super-K")
is another indirect detection experiment, located underground
in the Kamioka-Mozumi mine in Japan.  The detector consists of
50,000 tons of water and detects Cherenkov radiation from
incoming muons as well.  Super-K looks in the direction of the
sun, earth, and galactic center, and, like AMANDA, has not detected
any excess of muon rates above the expected background.\cite{SuperK}

\subsubsection{Antimatter}
Antimatter can be a excellent signal of WIMP annihilation precisely because
antimatter is relatively rare cosmically, and many of the astrophysical processes which create antimatter are well understood.  For example, the annihilation of WIMPs can also produce antiprotons
via $\chi \chi
\rightarrow q \bar{q}$ through hadronization (where the
dominate annihilation process yields $b$ quarks and
antiquarks), and positrons through secondary products of the
annihilation such as $W^+ W^-~(\mbox{and } ZZ),$
where $W (\mbox{or } Z) \rightarrow e^+ v_e$.  Unlike
gamma-rays and neutrinos, these products are charged and
thus affected by magnetic fields within space and also
lose energy due to inverse Compton and synchrotron
processes, so we cannot make any conclusions about where
the annihilations occurred.  We therefore study the flux
of antimatter particles from the galactic halo as a whole,
rather than assumed dense areas such as the galactic center
or large bodies.

Experiments searching for antimatter must be located near
the top of the earth's atmosphere; various other cosmic
rays and their consequential particle showers create too
large and uncertain of a background to make conclusive
analyses.  It is important, however, to still consider
and subtract any background caused by cosmic rays that
reach the edges of our atmosphere.  In 1994, the HEAT
Collaboration detected an excess of cosmic ray positrons
of energies around 10 GeV possibly caused by neutralino
self-annihilation, and confirmed this signal again in
2000.\cite{HEAT,HEAT2}  A ``boost factor," however, must
be applied to the WIMP annihilation rate of a smooth
halo in order to match the HEAT data; this is perhaps an
indication that we exist within an extremely clumpy halo,
or that there are other unknown sources of antimatter.
The Balloon-borne Experiment with Superconducting Spectrometer
(BESS) also detected antiprotons with energies up to 4 GeV
during its nine flights over several years.\cite{BESS}

Quite recently, the results from the PAMELA (a Payload for Antimatter
Matter Exploration and Light-nuclei Astrophysics) satellite-borne
experiment's flight from July 2006-February 2008 were released.
The collaboration found that the positron fraction increases
sharply over much of the range of 1.5-100 GeV and thus concluded
that a primary source, either an astrophysical object or
dark matter annihilation, must be present to account for the
abundance of cosmic-ray positrons.\cite{PAMELA} The data from PAMELA
also require heavy WIMP candidates or large boost factors associated
with non-uniform clumps in the dark matter distribution, thus
constraining the nature of the possible dark matter.  ATIC (Advanced 
Thin Ionization Calorimeter), a balloon-based experiment, also 
reported an excess of $e^-$ or $e^+$ at 300-800 GeV.  However, recent 
results from Fermi\cite{FERMI} and HESS\cite{HESS} (the High Energy 
Stereoscopic System, a set of four Cherenkov telescopes in Namibia) 
do not see the same electron-positron excess of ATIC leaving the issue 
far from settled (however, Fermi does see an excess similar to that seen by PAMELA).  
Further data is necessary to determine if excess
gamma ray and antimatter fluxes are indeed signals of dark matter 
annihilation or signatures of local astrophysical objects and backgrounds.

\section{Conclusion and Challenges}

The astrophysical and cosmological evidence for dark matter is both impressive and compelling.  What is perhaps the most striking are the multiple lines of evidence which point to the need for dark matter.  Elemental abundances from Big Bang Nucleosynthesis and fundamental anisotropies in the  Cosmic Microwave Background Radiation both predict very similar baryon (ordinary matter) abundances, yet each describes a completely separate era in the history of the universe in which very different physical processes are occurring.  Dark matter is necessary to both describe galaxies and clusters of galaxies, and is a necessary ingredient in the formation of large scale structure.  It is this concordance of evidence that makes dark matter more than just a ``fudge-factor"; although strange and unexpected, dark matter seems to be a fundamental and necessary component of our universe.

Although the composition and nature of dark matter is still unknown, theories like Supersymmetry or Kaluza-Klein theories of extra dimensions provide solid frameworks for attempting to understand dark matter.  Of all of the particle candidates for dark matter, perhaps the best motivated is the neutralino.  It is a typical WIMP: electrically neutral, weakly interacting, and massive, and through statistical mechanics in the early universe we can calculate abundances for the neutralino today which are consistent with it acting as the dark matter.  Other exotic candidates for dark matter exist from axions to Q-balls to WIMPzillas.  However outlandish the candidate, the hunt for dark matter continues.  The Large Hadron Collider at CERN will begin collisions at 3.5 TeV per beam in 2009-2010, ramping up to 7 TeV per beam most likely in 2011 and beyond, and will search for indications of supersymmetry and dark matter.  Indirect searches continue to hunt for gamma rays and antimatter which might provide evidence for dark matter; the current controversy between PAMELA and ATIC and FERMI and HESS results demonstrate the advances and challenges in indirect detection.  And finally, direct detection experiments continue to set more stringent limits on neutralino and WIMP scattering cross sections; these limits, as new technology is applied, are set to improve dramatically in the next decade with experiments like Super CDMS, GENIUS, and ZEPLIN IV.

Dark matter, of course, is not completely understood and faces challenges.  The primary challenge is that it remains undetected in the laboratory.  Another tension for dark matter is that it seems to possess too much power on small scales ($\sim$ 1 - 1000 kpc).  Numerical simulations of the formation of dark matter halos were performed by Klypin \emph{et al.} and show that, to explain the average velocity dispersions for the Milky Way and Andromeda, there should be five times more dark matter satellites (dwarf galaxies with a very small ordinary matter content) with circular velocity $>$ 10-20 km/s and mass $> 3 \times 10^8 ~ M_\odot$ within a 570 kpc radius than have been detected.\cite{Klypin}  In other words, although dark matter is crucial in forming structure, current models form \emph{too much} structure.  Another study, from B. Moore \emph{et al.}, shows that dark matter models produce more steeply rising rotation curves than we see in many low surface brightness galaxies, again suggesting that simulations produce an overabundance of dark matter.\cite{Moore}  Of course, discrepancies on small scales may be entirely due to astrophysical processes;  for example, photo-heating during reionization and/or supernova feedback particularly affect dwarf galaxies\cite{dwarf1,dwarf2}.  Although important to consider, these challenges faced by dark matter are dwarfed by the compelling evidence for the necessity of dark matter along with its successes in explaining our universe.  What makes this field so rich and vibrant is that work and research continue, and these challenges will lead to deeper understanding in the future.

Dark matter is an opportunity to learn more about the fundamental order of the universe.   Dark matter provides a tantalizing glimpse beyond the highly successful Standard Model of particle physics.  The discovery of neutralinos would prove the validity of supersymmetry and help bridge the ``desert" between the electroweak and the Planck scales.  But ultimately, we look at dark matter as a
mystery, one which will hopefully inspire physics and astronomy students in and out of the classroom.  As Einstein said, ``The most beautiful thing we can experience is the mysterious. It is the source of all true art and all science."

\section{Acknowledgements}

K.G. and G.D. would like to acknowledge support from the NASA Nebraska Space Consortium through grant NNG05GJ03H.

\section{Appendix: MOND Theories as Alternatives to Dark Matter}

One alternative to dark matter, particularly as an explanation for the 
non-Keplerian motions of rotating bodies, is called MOND (MOdified 
Newtonian Dynamics).  In 1983 M. Milgrom proposed that  the flat rotation curves observed in many galaxies may be explained without postulating any sort of  missing mass in the universe.\cite{Milgrom}
He instead introduced an acceleration constant to modify Newton's 
second law, which would at small accelerations account for the 
radius independent nature of stellar motion.

Rather than the usual $\vec{F} = m\vec{a}$, the equation at the 
heart of MOND is

\begin{equation}
\vec{F} = m \mu \Big(\frac{a}{a_0}\Big) \vec{a},
\end{equation}

\begin{displaymath}
\mu(x \gg 1) \approx 1, ~~~ \mu(x \ll 1) \approx x,
\end{displaymath}

\noindent where $\vec{F}$ is the force acting on an object of 
mass $m$ and acceleration $a = |\vec{a}|$, and $a_0 \approx 2 
\times 10^{-8} \mbox{ cm s}^{-2}$ is the acceleration constant 
determined by Milgrom (many other MOND theories have emerged 
with differing values for $a_0$).  For accelerations greater 
than or equal to $a_0$ (most accelerations we see in everyday 
life, including the motions of planets within our solar system), 
$x \approx 1$, and Newtonian dynamics can be used as usual.  
However, for very small accelerations such as for the orbits of 
objects far away from the galactic center, $a_0$ becomes 
significant; this is how MOND predicts and explains the flat 
rotation curves.

To demonstrate how MOND can explain flat rotation curves, we 
first consider the expression for the force of gravity $\vec{F}$ 
on a star and Milgrom's modification of Newton's second law:

\begin{equation}
\vec{F} = \frac{GMm}{r^2} = m \mu \Big(\frac{a}{a_0}\Big) \vec{a},
\end{equation}

\noindent where $G$ is the gravitational constant, $m$ and 
$M$ are the masses of the star and galaxy respectively, and 
$r$ is the radius of the star's orbit.  If we cancel $m$ from 
both sides and assume that $\mu (a/a_0) = (a/a_0)$ at a very 
large $r$, we are left with

\begin{equation}
\frac{GM}{r^2} = \frac{a^2}{a_0}.
\end{equation}

\noindent Solving for $a$ and using the relationship of 
acceleration with velocity and radius ($a=v^2/r$), we find

\begin{equation}
a = \frac{\sqrt{GMa_0}}{r} = \frac{v^2}{r}, ~~~ \mbox{and therefore,}
\end{equation}

\begin{equation}
v = \sqrt[4]{GMa_0},
\end{equation}

\noindent where $v$ has no dependence on $r$.  This relation 
has allowed various studies to use MOND to fit flat rotation 
curves quite successfully for several low and high surface 
brightness galaxies (LSB and HSB galaxies, respectively) based 
on luminous mass alone.\cite{Sanders,Sanders2,Brownstein}  
As MOND predicts, LSB galaxies show a larger departure from 
Newtonian dynamics where HSB galaxies show discrepancies only 
in their outer regions where gravitational attraction is 
considerably smaller.  MOND and TeVeS (the MONDian version of 
General Relativity)\cite{Bekenstein} have had success in 
predicting and describing other observed galactic dynamics 
as well.  For a recent review, see R. H. Sanders.\cite{Sanders3}

Despite these successes, MOND faces several major and critical challenges it has not been able to overcome. For example, when considering galaxy clusters, MOND  cannot account for density and temperature profiles and requires unseen matter.\cite{Aguirre}  Evidence for dark matter exists on many distance scales and MOND essentially only works on galactic scales.  Also, extremely low acceleration 
experiments (below $a_0$) have been conducted, finding no 
departure from Newton's second law and thus constraining MOND to
reduce to Newton's second law in laboratory conditions.\cite{Abramovici,Gundlach}
And finally, gravitational lensing evidence such as in the Bullet Cluster show that, in effect, the gravitational force points back not towards regular, observed baryonic matter but rather some form of
dark matter which is not observed optically.  MOND theories in their current forms cannot account for such a discrepancy easily, although more recent theories which wed MOND with a sterile neutrino are being developed.\cite{Angus}  However, for the above reasons and others, we feel that dark matter is a more promising solution to the puzzle of missing mass in the universe.

\end{document}